# Mixed Convection Heat Transfer and Flow of $Al_2O_3$-Water Nanofluid in a Square Enclosure with Heated Obstacles and Varied Boundary Conditions

Hashnayne Ahmed [a, b] *, Chinmayee Podder [a]

[a] *Department of Mathematics, Faculty of Science & Engineering, University of Barishal, Barishal 8200, Bangladesh*
[b] *Department of Mechanical and Aerospace Engineering, University of Florida, Gainesville, Florida 32611, USA*

**ABSTRACT**

*This paper studies the effects of mixed convection fluid motion and heat transmission of $Al_2O_3$ −water nanofluid in a square enclosure including two heated obstacles, with temperature and nanoparticle concentration being determined by the thermal conductivity and effective viscosity. The parametric observations of Richardson number, Reynolds number, cylinder rotating speed, and cavity inclination angles are investigated in the range of $0.1 \leq Ri \leq 10$, $1 \leq Re \leq 125$, $1 \leq \omega \leq 25$, and $0° \leq \gamma \leq 60°$ respectively on the thermal environment and flow arrangement inside the cavitation field. Adding nanoparticles to the base fluid enhances the heat transfer rate for both obstacles and all ranges of the parameters. The influence of wavy walls, changes in the nanofluid, and distinct positional effects also impact the flow characteristics and heat transfer process.*

List of symbols

| | |
|---|---|
| $c$ | Empirical constant for thermal boundary layer |
| $C_p$ | Specific heat at constant pressure ($JKg^{-1}K^{-1}$) |
| $d_p$ | Nanoparticle diameter ($nm$) |
| $D_f$ | Fractal dimension of the nanoparticles |
| $h$ | Heat transfer coefficient ($Wm^{-2}K^{-1}$) |
| $k$ | Thermal conductivity ($Wm^{-1}K^{-1}$) |
| $L$ | Length and height of the enclosure ($m$) |
| $Nu$ | Nusselt Number ($hx/k$) |
| $p$ | Pressure ($Pa$) |
| $P$ | Dimensionless pressure ($p/\rho_{nf}U_0^2$) |
| $Pr$ | Prandtl number ($\nu_f/\alpha_f$) |
| $R$ | Volume fraction of the nanoparticles |
| $Ra$ | Rayleigh number ($g\beta_f L^3 \Delta T/\nu_f \alpha_f$) |
| $Re$ | Reynolds number ($U_0 L/\nu_f$) |
| $T$ | Temperature ($K$) |
| $u, v$ | Velocity components in cartesian coordinates ($ms^{-1}$) |
| $U, V$ | Dimensionless velocity components |
| $U_0$ | Lid-driven velocity ($ms^{-1}$) |
| $x, y$ | Cartesian coordinates ($m$) |
| $X, Y$ | Dimensionless coordinates |

Greek symbols

| | |
|---|---|
| $\alpha$ | Thermal diffusivity ($m^2 s^{-1}$) |
| $\beta$ | Thermal expansion coefficient ($K^{-1}$) |
| $\gamma$ | Cavity inclination angle |
| $\varepsilon, \eta$ | Empirical constants for viscosity |
| $\theta$ | Dimensionless temperature ($T - T_c/T_h - T_c$) |
| $\rho$ | Density ($Kgm^{-3}$) |
| $\phi$ | Concentration ratio of the nanoparticles |
| $\omega$ | Cylinder rotating speed ($s^{-1}$) |

Subscripts

| | |
|---|---|
| $ave$ | Average |
| $c$ | Cold |
| $eff$ | Effective |
| $f$ | Base fluid |
| $h$ | Hot |
| $max$ | Maximum |
| $min$ | Minimum |
| $nf$ | Nanofluid |
| $s$ | Solid particle |
| $w$ | Wall |

## Introduction

Different engineering and geophysical systems are very interested in the phenomena known as mixed convection heat transfer, which results from the interaction of forced and natural convection. The study of mixed convection in these confined areas is of great relevance since fluid-filled enclosures are essential parts of such systems. While mixed convection boundary layers outside enclosures have been extensively researched, the flow and heat transfer inside a hollow display unique properties because of the intricate interactions between the finite-size fluid system and the surrounding walls. The study of mixed convection in enclosures is a difficult but crucial topic of research because this complexity gives conception to various flow patterns and heat transport processes.

The introduction of heated barriers into a vented cavity leads to an intriguing circumstance. In these circumstances, the cavity's motion of the lid generates a shear force that interacts with the buoyancy force resulting from temperature fluctuations within the hollow, causing fluid flow and heat transfer. Under mixed convection circumstances, the interaction between shear-driven and buoyancy-driven flows in a small cavity becomes extremely complex. Therefore, it is crucial to have a thorough understanding of how fluid flows and how heat transfers in such lid-driven cavities.

Cooling systems provide substantial difficulties in a variety of industrial applications and locations where effective heat transmission is essential. Larger heat transfer surfaces are frequently required by conventional heat transfer methods, increasing the size of the device unintentionally. Furthermore, the heat conductivity of common fluids such as water, ethylene glycol, and mineral oils is restricted [5]. To overcome these drawbacks, nanofluids have shown promise as a means of improving the effectiveness of heat transmission in these systems. The improvement of heat transfer rates in several real-world engineering applications has been demonstrated by nanofluids, which are made up of nano-sized solid particles floating in a base fluid with better thermal conductivity.

Numerous studies have been done to investigate the characteristics and uses of nanofluids in confined spaces. To understand the complex heat transmission behavior of $TiO_2$ − water nanofluid when it is heated from the bottom inside of a restricted space, Wen and Ding embarked on an in-depth study. They calculated the viscosity of the nanofluid and its effective heat conductivity, using a variety of cutting-edge models for thorough investigation [1]. In $Al_2O_3$ −water nanofluids, Hwang et al. demonstrated that lowering average temperature, raising volume fraction, or reducing nanoparticle size all enhanced stability and heat transfer in a heated rectangular enclosure [2]. In a 2D hollow containing a water-copper nanofluid, Khanafer et al. numerically investigated natural convection. Increased heat transfer across all Grashof values was correlated with higher nanoparticle volume percentage [13]. Jou and Tzeng's numerical analysis of differentially heated rectangular cavities filled with a nanofluid revealed consistent results. Increasing buoyancy parameter and nanofluid volume fraction raised the average heat transfer coefficient [12]. The effect of a square cavity's inclination angle on the natural convection of Cu-water nanofluid was studied by Abu-nada and Oztop [9]. According to their research, the inclination angle can be employed to regulate heat and fluid transmission. Additionally, the influence of the inclination angle on the percentage of heat transfer augmentation diminished at lower Rayleigh numbers [8]. Ogut carried out a computational analysis of natural convection heat transfer in a square cavity with an inclination filled with water-based nanofluids. The vertical sides were adiabatically heated on the left, cooled on the right, and heated continuously on the left. They discovered that as particle volume fractions and Rayleigh numbers rose, so did the average rate of heat transfer [19].

Researchers have studied how nanofluids behave in terms of heat transmission in cavities with varying geometries and boundary conditions. Particle size, volume fraction of nanoparticles, and other factors have all been studied in relation to fluid flow patterns and heat transmission rates. The intricacy of heat transfer processes is further increased by the inclusion of heated impediments inside the cavities, and these situations have direct technical applications in solar collectors, heat exchangers, thermal insulation, and the cooling of electronic equipment and chips using nanofluids. In a cavity that was variably heated and had an internal obstruction, Mezrhab et al. quantitatively examined the interactions between radiation and natural convection. [20]. The temperature in the square cavity is homogenized by radiation exchange, which raises the average Nusselt number. Chen discovered that the Rayleigh number, eccentricity, and geometric configurations all affect the heat and fluid flow patterns in the annulus when natural convection is used to transfer heat from the inside heated sphere to the outer vertically eccentric cold cylinder [21]. Liquid gallium was placed inside a horizontal circular cylinder with an interior coaxial triangle cylinder, and Yu et al. mathematically modeled the transient natural convection heat transfer of the material [4]. They found that placing the top side of the inner triangular cylinder horizontally improved the time averaged Nusselt number for Grashof numbers greater than $10^5$.

Due to its distinct mixture of natural and forced convection, mixed convection stands out among the numerous heat transfer techniques used in these applications. The phenomena of mixed convection with lid-driven effects have several applications, including the cooling of electronic devices, lubricating technologies, drying procedures, and equipment used in chemical processing. Due to the interaction of buoyancy and shear forces, mixed convection heat transfer is an intricate procedure [22]. Recent research employing nanofluids to study mixed convection heat transport has highlighted the importance of this phenomena in real-world engineering settings.

In a study by Tiwari and Das, they investigated heat transfer in a square chamber with a two-sided lid and a specific fluid termed $Cu$–water nanofluid. The cavity included insulated top and bottom walls as well as heated movable sidewalls. They found that the Richardson number and the orientation of the moving walls both had an impact on how the fluid moved and transferred heat inside the cavity. Interestingly, increasing the number of nanoparticles in the nanofluid improved heat transport when the Richardson number was equal to one, as demonstrated by larger average Nusselt numbers. This indicates that the heat transfer properties of the nanofluid were enhanced by a larger nanoparticle concentration [47]. Muthtamilselvan et al. conducted different research in which heat transmission was also examined using Cu-water nanofluid in a lid-driven rectangular enclosure with mixed convection flow. Both research shed critical light on how nanofluids might affect heat transport in closed environments, which may have significant applications in a variety of industries [34]. According to the study, linear changes in the average Nusselt number for stable solid volume are associated with fluid flow and heat transmission in rectangular cavities depending on the aspect ratio and solid volume percentage. Using the finite volume technique and SIMPLER method, Guo and Sharif examined mixed convection in rectangular cavities with sliding isothermal sidewalls and constant heat flux [7]. Their research revealed that moving the heat source closer to the sidewalls improved the average Nusselt number. Shahi et al. performed a numerical analysis of mixed convection in a vented, partially heated from below square cavity in a separate experiment [35]. According to their observations, the average Nusselt number was improved by moving the heat source closer to the sidewalls. Malik et al. and Esfe et al. also investigated mixed convection in a lid-driven cavity filled with nanofluid and according to their findings, $Ri$ is mostly affecting the heat transfer rate [48, 51].

The effective viscosity and thermal conductivity of nanofluids, which are essential for predicting heat transport within enclosures, cannot be predicted with sufficient accuracy by current classical models [36]. To predict these features more accurately for nanofluids, researchers have developed innovative models. Numerous research has looked at how different properties, such viscosity and thermal conductivity, affect natural convection in cavities filled with nanofluids. Ho et al. highlighted the significance of the models they developed for the viscosity and thermal conductivity of the nanofluid in estimating heat transfer augmentation or reduction within the enclosure relative to the base fluid [37]. The effects of inlet and outlet positioning on the mixed convection of a nanofluid in a square cavity were investigated by Mahmoudi et al. [38]. The impact of altering thermal conductivity and viscosity on natural convection in cavities filled with $Al_2O_3$ −water and $CuO$ −water nanofluids was studied by Abu-Nada et al. Their research showed that the viscosity models had a greater influence on the average Nusselt number than the thermal conductivity models did for large Rayleigh numbers [9]. Mahmoodi also employed $Al_2O_3$ −water nanofluid to simulate mixed convection in rectangular chambers with a heated, sliding bottom lid and cold, right left, and top walls [39]. The effects of nanofluid variable features on mixed convection in a square cavity were investigated by Mazrouei Sebdani et al. [40].

In a square enclosure with two cold walls (left and right), a cold horizontal top wall, and a sliding hot bottom wall, Arefmanesh and Mahmoodi numerically examined the impact of dynamic viscosity models for an $Al_2O_3$ −water nanofluid on mixed convection [39]. The investigation found that increasing the solid volume percentage of nanoparticles resulted in higher average Nusselt numbers on the heated wall of the enclosure. Nasiri et al. have investigated a square cavity with a spinning cylinder and phase-shifting nanofluids within it [45]. In a wavy lid-driven porous enclosure filled with CNT-water nanofluid under the influence of a magnetic field, Hamzah et al. investigated mixed convection with entropy generation [46].

Addressing these research gaps, this article presents a numerical investigation of mixed convection heat transfer in a square double lid-driven cavity filled with $Al_2O_3$ −water nanofluid. Using the models suggested by Jang et al. and Abu-Nada et al., the dynamic viscosity in the research was computed [9, 44]. The Xu model was used in the current experiment to calculate the thermal conductivity of the nanofluid [43]. To calculate the thermal conductivity and dynamic viscosity of the nanofluid in this study, models based on parameters such as nanoparticle diameter, diameter ratio, temperature, and nanoparticle volume percentage were used. In an enclosure with a nanofluid intake and exit, the study investigates the heat conduction and fluid flow characteristics of mixed convection. Because it presents unique models with adjustable properties, this research differs from earlier numerical studies in the subject. The study further investigates the impact of position variation, wavy wall, and rotating speed of a circular barrier on heat transfer, temperature, and flow fields inside the enclosure, which also includes the presence of two heated obstacles (a square block and a circular cylinder). By exploring these objectives, our research hopes to greatly advance our understanding of mixed convection heat transfer and flow in enclosures containing heated barriers and nanofluids, with consequences for real-world engineering applications.

## Mathematical description of the problem

The following Fig. 1 depicts a two-dimensional square cavity with physical dimensions ($L$) that was considered for this investigation. External cold nanofluid enters the cavity at the bottom of the left insulated wall (inlet, $0.2L$) and exits through the top of the right insulated wall (outlet, $0.2L$). A square block of length $0.2L$ and a revolving cylinder of diameter $0.2L$ are considered in the square cavity's middle. For the sake of physics, gravity's acceleration acts against the down wall.

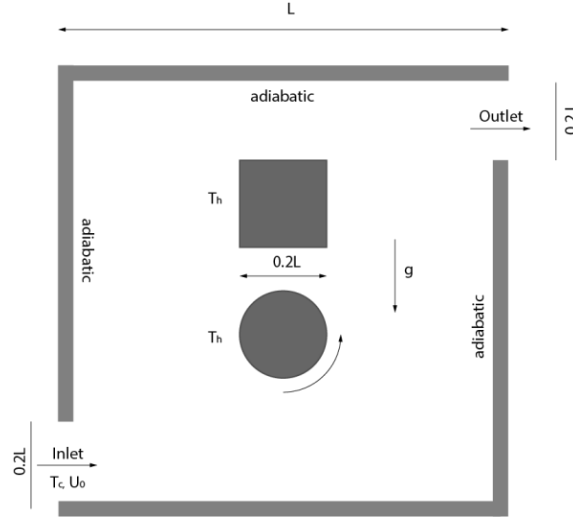

**Fig. 1.** Schematic diagram of the physical problem.

Because the cavity's length perpendicular to its plane is presumed to be sufficient, the investigation is classified as two-dimensional. A suspension of $Al_2O_3$ nanoparticles (with a particle diameter of $47\ nm$) in water are placed in the cavity, and there is no slide between them. Nanoparticles and the base fluid are in thermal equilibrium, and the nanofluid is believed to be incompressible. The following Table 1 shows the thermophysical characteristics of nanoparticles with water as the base fluid at $T = 25°C$.

**Table 1.**
Thermo-physical properties of water and nanoparticles ($Al_2O_3$) at T = $25^0$C.

| Physical Property | Water, $H_2O$ | Nanoparticle, $Al_2O_3$ |
|---|---|---|
| $C_p\ (J/Kg.K)$ | 4179 | 765 |
| $\rho\ (Kg/m^3)$ | 997.1 | 3970 |
| $K\ (W.m^{-1}.K^{-1})$ | 0.6 | 25 |
| $\beta\ (1/K)$ | $21 \times 10^5$ | $0.85 \times 10^{-5}$ |
| $\mu\ (Kg/m.s)$ | $8.9 \times 10^4$ | – |
| $Diameter\ (m)$ | – | $47 \times 10^{-9}$ |

The governing equations for a steady, two-dimensional laminar and incompressible flow are expressed by:

$$\nabla u = 0 \tag{1}$$

$$u.\nabla u = -\frac{1}{\rho_{nf}}\nabla p + \nu_{nf}\nabla^2 u + g\beta_{nf}(T - T_c) \tag{2}$$

$$u.\nabla T = \alpha_{nf}\nabla^2 T \tag{3}$$

The dimensionless form of the above governing equations can be obtained using the following parameters:

$$X = \frac{x}{L}, Y = \frac{y}{L}, U = \frac{u}{U_0}, V = \frac{v}{U_0},$$
$$\Delta T = T_h - T_c, \theta = \frac{T - T_c}{\Delta T}, \text{and } P = \frac{p}{\rho_{nf}U_0^2} \tag{4}$$

The two-dimensional dimensionless equations for the conservation of total mass, momentum, and energy of the nanofluid can be written as:

$$\frac{\partial U}{\partial X} + \frac{\partial V}{\partial Y} = 0 \tag{5}$$

$$U\frac{\partial U}{\partial X} + V\frac{\partial U}{\partial Y} = -\frac{\partial P}{\partial X} + \frac{\nu_{nf}}{\nu_f} \cdot \frac{1}{Re}\left(\frac{\partial^2 U}{\partial X^2} + \frac{\partial^2 U}{\partial Y^2}\right) \tag{6}$$

$$U\frac{\partial V}{\partial X} + V\frac{\partial V}{\partial Y} = -\frac{\partial P}{\partial X} + \frac{\nu_{nf}}{\nu_f} \cdot \frac{1}{Re}\left(\frac{\partial^2 V}{\partial X^2} + \frac{\partial^2 V}{\partial Y^2}\right) + \frac{\beta_{nf}}{\beta_f} \cdot Ri \cdot \theta \tag{7}$$

$$U\frac{\partial \theta}{\partial X} + V\frac{\partial \theta}{\partial Y} = \frac{\alpha_{nf}}{\alpha_f} \cdot \frac{1}{Pr} \cdot \frac{1}{Re}\left(\frac{\partial^2 \theta}{\partial X^2} + \frac{\partial^2 \theta}{\partial Y^2}\right) \tag{8}$$

where,

$$Re = \frac{U_0 L}{\nu_f}, Pr = \frac{\nu_f}{\alpha_f}, Ri = \frac{g\beta_f \Delta T L}{U_0^2} = \frac{Ra}{Pr \cdot Re^2} \tag{9}$$

The boundary conditions are:
$u = v = 0$, and $\partial T/\partial y = 0$ on the top wall.
$u = v = 0$, and $\partial T/\partial y = 0$ on the bottom wall.
$u = v = 0$, and $\partial T/\partial x = 0$ on the right wall. (10)
$u = v = 0$, and $\partial T/\partial x = 0$ on the left wall.
$u = v = 0$, and $T = T_h$ on the rectangular block boundaries.
$u = v = 0$, and $T = T_h$ on the circular cylinder boundaries.

The dimensionless boundary conditions become,
$U = V = 0$, and $\partial \theta/\partial Y = 0$ on the top wall.
$U = V = 0$, and $\partial \theta/\partial Y = 0$ on the bottom wall.
$U = V = 0$, and $\partial \theta/\partial X = 0$ on the right wall. (11)
$U = V = 0$, and $\partial \theta/\partial X = 0$ on the left wall.
$U = V = 0$, and $\theta = 1$ on the rectangular block boundaries.
$U = V = 0$, and $\theta = 1$ on the circular cylinder boundaries.

Also, by comparing the dimensional and non-dimensional forms, $\rho_{nf} = 1$.

The effective viscosity of nanofluid was computed by the following Eq. 12:

$$\mu_{eff} = \mu_f(1 + 2.5\phi)\left[1 + \eta\left(\frac{d_p}{L}\right)^{-2\varepsilon} \phi^{\frac{2}{3}}(\epsilon + 1)\right] \tag{12}$$

A verified model for spherical particle nanofluids was published by Jang et al. [34]. The viscosity of water is determined by the following equation and contains the empirical constants ($-0.25$ and $280$ for $Al_2O_3$).

$$\mu_{H_2O} = (1.2723 \times T_{rc}^5 - 8.736 \times T_{rc}^4 + 33.708 \times T_{rc}^3 - 246.6 \times T_{rc}^2 \\ + 578.78 \times T_{rc} + 1153.9) \times 10^6 \tag{13}$$

Where $T_{rc} = \log(T - 273)$.

The effective density, heat capacitance, thermal expansion coefficient, and thermal diffusivity of the nanofluid is computed by the following equations:

$$\rho_{nf} = \phi \rho_s + (1 - \phi)\rho_f \tag{14}$$
$$(\rho C_p)_{nf} = \phi(\rho C_p)_s + (1 - \phi)(\rho C_p)_f \tag{15}$$
$$(\rho \beta)_{nf} = \phi(\rho \beta)_s + (1 - \phi)(\rho \beta)_f \tag{16}$$

$$\alpha_{nf} = \frac{k_{nf}}{(\rho C_p)_{nf}} \tag{17}$$

The model presented by Hamilton and Crosser is used to calculate the effective thermal conductivity of nanoparticles in liquid as stationary (the $H-C$ model) [35]:

$$\frac{k_{stationary}}{k_f} = \frac{k_s + 2k_f - 2\phi(k_f - k_s)}{k_s + 2k_f + \phi(k_f - k_s)} \tag{18}$$

The following model was proposed by Xu et al., and it has been chosen in this study to describe the thermal conductivity of nanofluids [33]:

$$\frac{k_{nf}}{k_f} = \frac{k_{stationary}}{k_f} + \frac{k_c}{k_f} = \frac{k_s + 2k_f - 2\phi(k_f - k_s)}{k_s + 2k_f + \phi(k_f - k_s)} \\ + c \frac{Nu_p \, d_f (2 - D_f) D_f}{Pr \, (1 - D_f)^2} \frac{\left[\left(\frac{d_{max}}{d_{min}}\right)^{1-D_f} - 1\right]^2}{\left(\frac{d_{max}}{d_{min}}\right)^{2-D_f} - 1} \frac{1}{d_p} \tag{19}$$

The heat convection brought on by Brownian motion is represented by the second term in the above equation. The fluid under study and the type of nanoparticle under study are both factors that affect the empirical constant, $c$. In this study, the fluid molecular diameter of water, $d_f$ is set at $4.5 \times 10^{-10}$ m, and the Nusselt number for liquid flow around a spherical particle, $Nu_p$ is assumed to be 2. Also, $d_p$ and $\phi$ are the mean nanoparticle diameter and volume percentage of nanoparticles, respectively and $Pr$ is the Prandtl number. The fractal dimension $D_f$ is determined by,

$$D_f = 2 - \frac{\ln \phi}{\ln \left(\frac{d_{p,min}}{d_{p,max}}\right)} \tag{20}$$

Where $d_{p,max}$ and $d_{p,min}$ are the maximum and minimum diameters of nanoparticles, respectively. The ratio of minimum to maximum nanoparticles $d_{p,min}/d_{p,max}$ is $R$:

$$d_{p,max} = d_p \cdot \frac{D_f - 1}{D_f} \left(\frac{d_{p,min}}{d_{p,max}}\right)^{-1} \\ d_{p,min} = d_p \cdot \frac{D_f - 1}{D_f} \tag{21}$$

The effective thermophysical properties of the nanofluid in the above equations are calculated at room temperature, $T = 25^0 C$.

In this study, the software package COMSOL Multiphysics 5.5 [https://www.comsol.com/release/5.5] is utilized to model and simulate the fluid flow and heat transfer mechanisms in the square cavity problem. The modified Galerkin weighted residual Finite Element Technique is used for the explanation of the problem governing equations (Eq. 5 – Eq. 8) with boundary conditions (Eq. 11) and nanofluid properties of Table 1.

### Grid testing and validation

Numerous factors, including thermal conductivity, heat capacitance, viscosity, the structure of nanofluid flow, volume fraction, size, and fractal distributions of nanoparticles, influence the Nusselt number ($Nu$), resulting in notable local variations on hot walls [50].

$$Nu = -\left(\frac{k_{nf}}{k_f}\right)\left(\frac{\partial \theta}{\partial X}\right) \tag{22}$$

It is possible to get the average Nusselt number along the wall by integrating the local Nusselt number over it [49].

$$Nu_{ave} = \frac{1}{L}\int_0^L Nu\, dY \tag{23}$$

To ascertain if the mesh arrangement is sufficient and to verify that the findings are independent of the grid, several grid sensitivity tests were conducted using COMSOL's customized settings of element size (these sizes are even more dense than the COMSOL default setting options of finer, extra fine, and extremely fine.). As seen in the accompanying Fig. 3, triangular elements are employed. To get The Nusselt number, $Nu$ is calculated with the setting of $Re = 10, Pr = 6.96, Ri = 10, \phi = 0.06, d_p = 5nm$ and initial flow velocity of $1ms^{-1}$ with the cylinder rotation of $20ms^{-1}$ for each number of elements, as shown in Fig. 2. As can be observed, the maximum element size of $0.0108mm$ yields the required accuracy with the highest Nusselt number 11.15414 (For the rectangle, $Nu = 0.588615$ and for the circle, $Nu = 10.56552$) and was hence applied for all simulation exercises in this work, as presented in the following section.

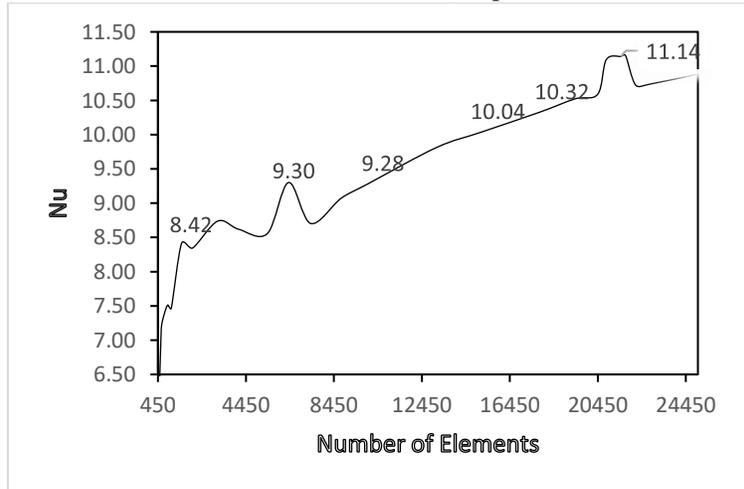

**Fig. 2.** Grid independence test.

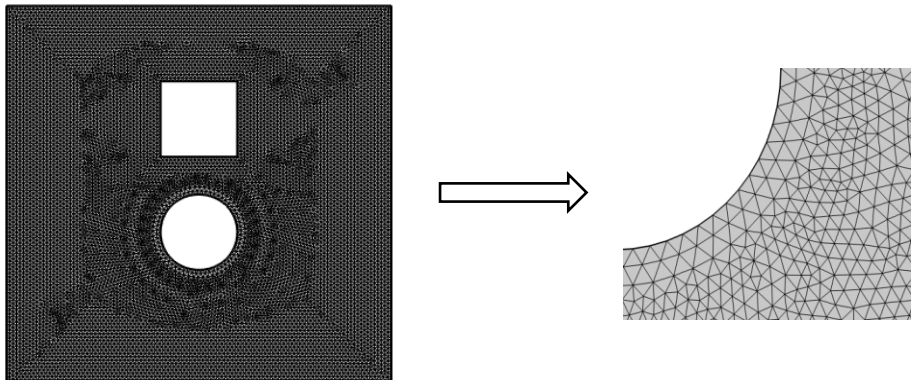

**Fig. 3.** Mesh structures with 21710 elements for this study.

We examined a square chamber with a heated left wall and a cooled right wall, while the top and bottom walls remained adiabatic, to confirm our computational setup. In this instance, there were no obstructions inside the cage. By comparing the maximum velocity along the $x$ and $y$ axes, and averaged Nusselt number along the hot wall of the square cavity obtained in this work with similar solutions [36], an excellent agreement was obtained as shown in Table 2.

Table 2.
Comparison of pure fluid solutions with previous works in a square cavity for Re = 1000 and Pr = 0.7 with different Rayleigh numbers.

| Index | Present Study | Hemmat Esfe et al. [3] | Lin & Violi [36] | Tiwari & Das [37] | Hadjiso-phocleous et al. [38] | Markatos and Pericleous [40] | Davis [41] |
|---|---|---|---|---|---|---|---|
| $Ra = 10^3$ | | | | | | | |
| $u_{max}$ | 3.6482 | 3.619 | 3.597 | 3.642 | 3.544 | 3.544 | 3.649 |
| $y$ | 0.8167 | 0.811 | 0.819 | 0.804 | 0.814 | 0.832 | 0.813 |
| $v_{max}$ | 3.6954 | 3.697 | 3.690 | 3.703 | 3.586 | 3.593 | 3.697 |
| $x$ | 0.1746 | 0.180 | 0.181 | 0.178 | 0.186 | 0.168 | 0.178 |
| $Nu_{ave}$ | 1.1178 | 1.114 | 1.118 | 1.087 | 1.141 | 1.108 | 1.118 |
| $Ra = 10^4$ | | | | | | | |
| $u_{max}$ | 16.1858 | 16.052 | 16.158 | 16.144 | 15.995 | 16.180 | 16.178 |
| $y$ | 0.8254 | 0.817 | 0.819 | 0.822 | 0.814 | 0.832 | 0.823 |
| $v_{max}$ | 19.6066 | 19.528 | 19.648 | 19.665 | 18.894 | 19.440 | 19.617 |
| $x$ | 0.1222 | 0.110 | 0.112 | 0.110 | 0.103 | 0.113 | 0.119 |
| $Nu_{ave}$ | 2.2442 | 2.215 | 2.243 | 2.195 | 2.290 | 2.201 | 2.243 |
| $Ra = 10^5$ | | | | | | | |
| $u_{max}$ | 34.8301 | 36.812 | 36.732 | 34.300 | 37.144 | 35.730 | 34.730 |
| $y$ | 0.8516 | 0.856 | 0.858 | 0.856 | 0.855 | 0.857 | 0.855 |
| $v_{max}$ | 68.3839 | 68.791 | 68.288 | 68.765 | 68.910 | 69.080 | 68.590 |
| $x$ | 0.0698 | 0.062 | 0.063 | 0.0594 | 0.061 | 0.067 | 0.066 |
| $Nu_{ave}$ | 4.5172 | 4.517 | 4.511 | 4.450 | 4.964 | 4.430 | 4.519 |
| $Ra = 10^6$ | | | | | | | |
| $u_{max}$ | 65.0775 | 66.445 | 66.470 | 65.587 | 66.420 | 68.810 | 64.630 |
| $y$ | 0.8516 | 0.873 | 0.869 | 0.839 | 0.897 | 0.872 | 0.850 |
| $v_{max}$ | 219.8262 | 221.748 | 222.340 | 219.736 | 226.400 | 221.800 | 217.360 |
| $x$ | 0.0349 | 0.040 | 0.038 | 0.042 | 0.021 | 0.038 | 0.038 |
| $Nu_{ave}$ | 8.8065 | 8.795 | 8.758 | 8.803 | 10.390 | 8.754 | 8.799 |

Indicating a considerable influence for small average nanoparticle sizes, Fig. 4 validates the Xu's model for effective thermal conductivity as a function of $R$ and $d_p$, two crucial factors [3].

As can be seen in Fig. 4, the ratio of thermal conductivity of nanofluid to the thermal conductivity of the fluid increases as the volume fraction of nanoparticles, $R$, increases and the highest value of $K_{nf}/K_f$ is obtained for $R = 0.06$. The highest ratio is obtained for $d_p = 5\ nm$ shown by the black line graph with the diamond mark, while the dash, dot-dash, and dot lines indicate the thermal conductivity ratios for $d_p = 25nm$, $d_p = 50nm$, and $d_p = 100nm$, respectively.

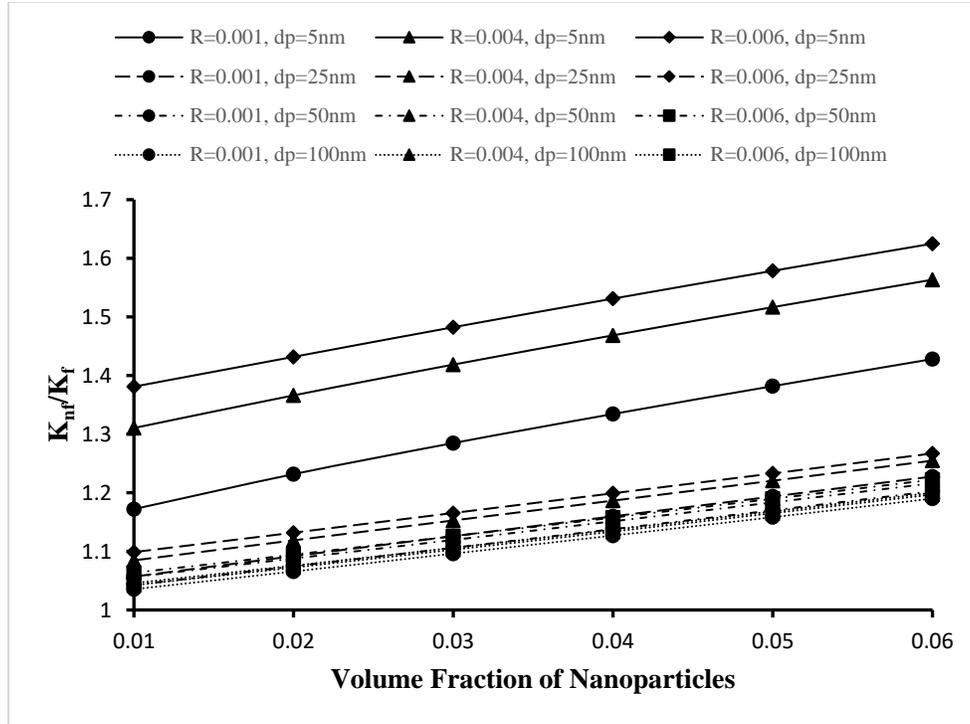

**Fig. 4.** Dimensionless effective thermal conductivity of $Al_2O_3$-water nanofluid against nanoparticle concentration, varying mean nanoparticle diameters and fractal distributions.

The ratio of thermal conductivity of the nanofluid drops as the nanoparticle diameter grows, whereas the ratio of thermal conductivity of the nanofluid increases as the volume fraction of nanoparticles increases. That is, as the volume fraction of nanoparticles grows and the diameter of the nanoparticles decreases, the thermal conductivity of the nanofluid increases.

## Results and discussion

In this chapter, heat transfer and flow characteristics inside a square cavity with a filled and inflow of $Al_2O_3$−water nanofluid has been studied. Inside the cavity, we considered a hot square obstacle and a hot cylinder rotating in the anti-clockwise direction. Influence on the streamlines and isotherms for some parameters including Richardson number, Reynolds number, cavity inclination angle, the rotational speed of the hot cylinder, changes in the nanofluid, change of position of the obstacles, and changes in the boundary conditions for the cavity and obstacles have been investigated. Also, the effects of the Richardson Number and Reynolds number on the local Nusselt number calculated at the hot walls have been plotted.

## Effects of Ri and Re on mixed convection

The effects of temperature gradient on the streamlines and isotherms inside the square cavity with $Al_2O_3$ −water nanofluid is shown in the figures below for different parameters. We know from the definition ($Ri = Gr/Re^2$) that when the Richardson number increases, the flow become more stable due to the influence of decreasing $Re$. When the Reynolds number rises, the flow velocity rises, the kinematic viscosity drops, and the laminar flow becomes turbulent. When $Re > 2800$, the flow patterns change from laminar to turbulent. In a similar manner, the viscous forces become stronger than the buoyancy forces due to the increasing $Ri$. On a general approximation of laminar flow in a square cavity, forced convection occurs for $Ri \ll 0.1$ (strong buoyancy force dominating over viscous force, highly unstable flow), mixed convection occurs for $0.1 \leq Ri \leq 10$ (balanced interaction between buoyancy and viscous forces, steady and smooth flow), and natural convection occurs for $Ri \gg 10$ (weak buoyancy force dominated by viscous force, stable flow).

Fig. 5 and Fig. 6 display the streamlines and isotherms respectively for different Richardson numbers at $\phi = 0.06, \omega = 20s^{-1}, R = 0.004, d_p = 5nm$ and at constant Reynolds number, $Re = 100$. At $Ri = 0.1$, with the dominance of the entrance fluid force over the buoyancy force, the streamlines have been expanded over the square obstacles, and consequently, velocity and pressure decreased. Moreover, four small and large vortexes have been formed around the rotating hot cylinder due to the forced convection. Also, it is clear from Fig. 5 that the streamlines are very intense near the downside of the rotating cylinder, which indicates a very high velocity of the fluid flow in these areas. At $Ri = 0.1$, the severe intensity of the isotherms close to the hot rotating cylinder is observed in Fig. 6. Here, the severe intensity of the isotherms means a high-temperature gradient and that indicates high heat transfer in the cavity in this position near the inlet. At $Ri = 1$, the effect of the buoyancy force decreases, and therefore, streamline expands and occupies a larger space in the cavity with the reduced number of vortices. The isotherms display a slight decrease in intensity close to the hot cylinder walls and a decrease in the temperature gradient. In this case, the heat transfer rate is smaller than that at $Ri = 0.1$. At $Ri = 3$, the streamlines around the hot square obstacle decrease mean the increase of velocity and pressure in the downsides, near the rotating cylinders. But after that, as the Richardson number is increased up to $Ri = 10$ from $Ri = 5$, the streamlines increase above the square obstacles, and thus, velocity and pressure decrease. Due to this, streamlining expands as well as the number of vortices remains constant. From Fig. 6, we observe that the isotherm line remains almost the same (a very slight decrease) means about the same temperature gradient around the hot rotating cylinder for $Ri = 5$ to $Ri = 10$.

Similarly, Fig. 7 and Fig. 8 display the streamlines and isotherms respectively for different Reynolds numbers at $\phi = 0.06, \omega = 20s^{-1}, R = 0.004, d_p = 5nm$ and at constant Richardson number, $Ri = 7$. At $Re = 1$, the streamline is expanded over the hot square obstacle as well as on the downside of the cavity indicating a decreased situation of velocity and pressure in the middle flow region. As the Reynolds number is increased up to $Re = 125$ from $Re = 1$, the streamline expansions remain almost the same except for changes in the number and size of the vortices. At $Re = 125$, five vortices are generated around the hot rotating cylinder and the streamlines also become more intense than the lower Reynolds number indicating that the flow in the middle is increased as well as there is more fluid flow near the boundaries with slower velocity and pressure than middle regions. From Fig. 8, we observe that the isotherm lines become more intense around the hot rotating cylinders for increased Reynolds numbers. At $Re = 1$, isotherms remain scattered around the cavity but as it increases, the severe intensity of isotherm line distribution around the hot rotating cylinder confirms the increased heat transfer validated by Fig. 17 (increased Nusselt number for higher Reynolds number).

Fig. 9 and Fig. 10 display the surface view of streamlines and isotherms respectively for the effects of Richardson number and Reynolds number shown in Fig. 5 to Fig. 8. From Fig. 9, we see that the inflow velocity ($1 \ ms^{-1}$) increases up to $2.5 \ ms^{-1}$ due to the rotational speed, ($\omega = 20$ rotation per second). The middle region (corner line along the inlet and outlet) holds the variational flow characteristics, and the boundary region moves with slower velocity on a common note. Fig. 10 demonstrates the temperature characteristics of the square cavity where the vast change of the temperature gradient happens due to the rotational speed of the cylinder. Heat transfer is more for the rotating cylinder than the square obstacle and the increased Reynolds number ensures more heat transfer than any other tests performed. The effects of the Richardson number are calculated with the constant Reynolds number, $Re = 100$, and the effects of the Reynolds number are calculated with the constant Richardson number, $Ri = 7$.

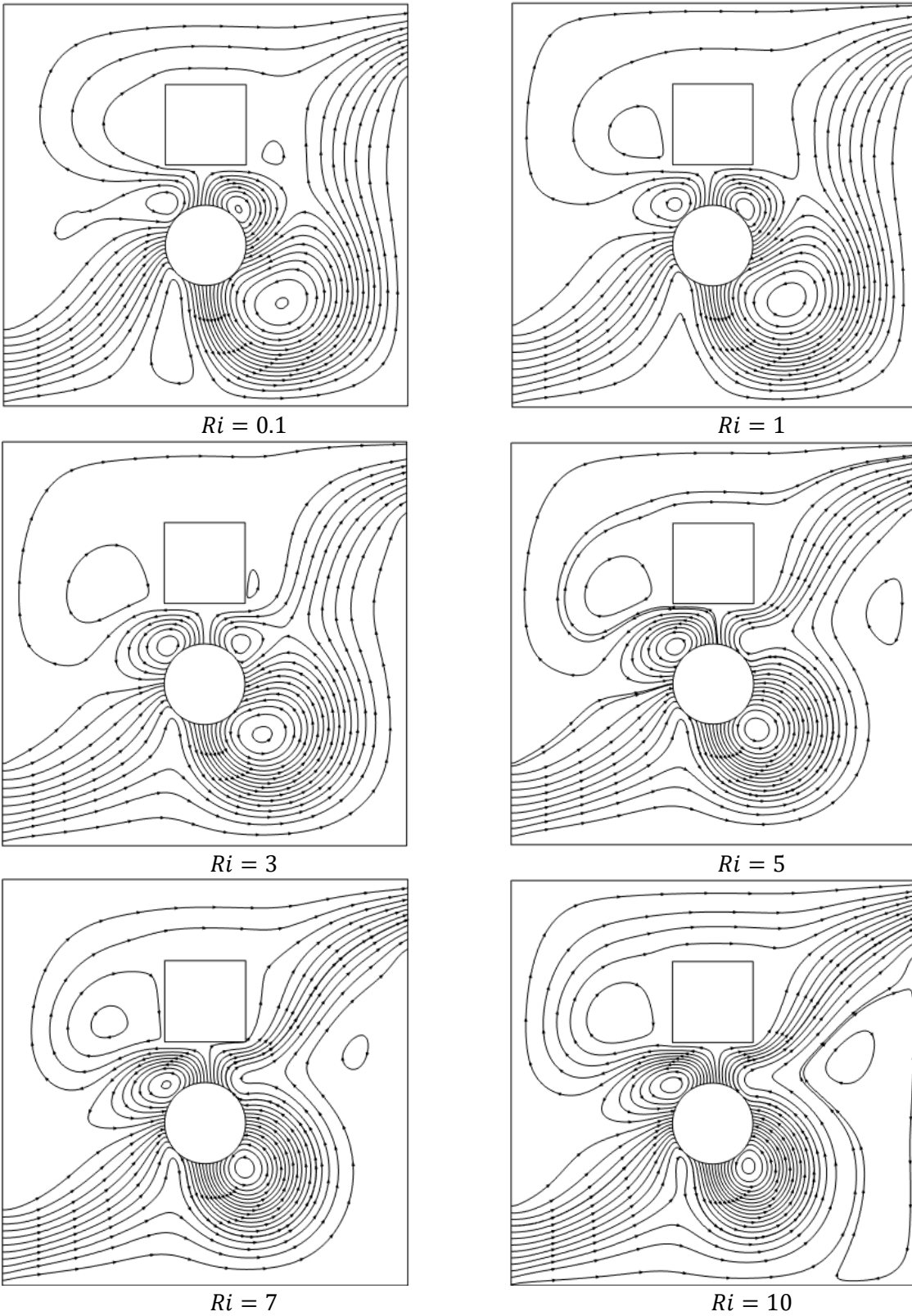

**Fig. 5.** Streamline variation for the variation in Richardson number (Ri) at $\phi = 0.06$, $\omega = 20s^{-1}$, $R = 0.004$, $d_p = 5$nm and Re = 100.

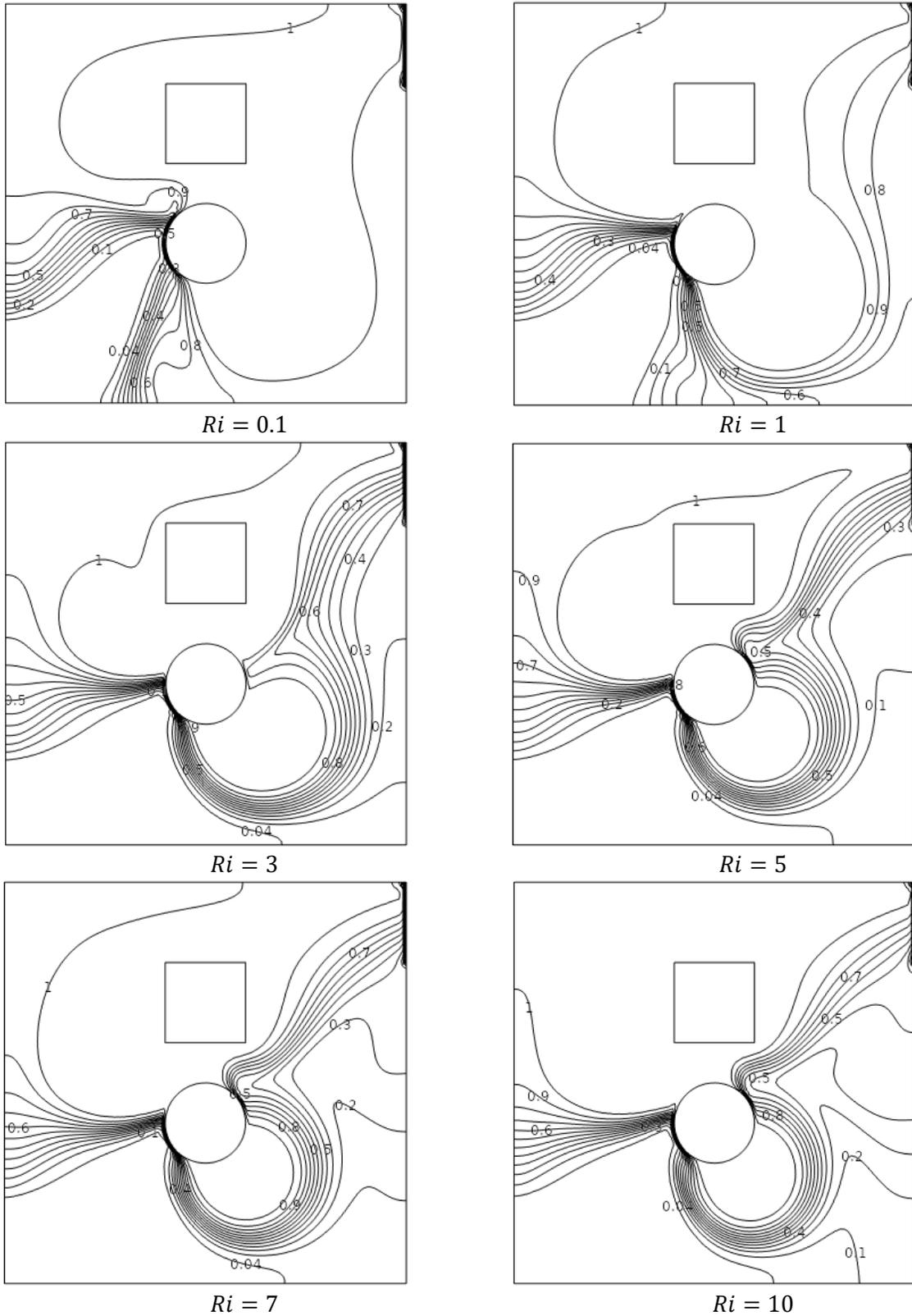

**Fig. 6.** Isothermal contours variation for the variation in Richardson number (Ri) at $\phi = 0.06, \omega = 20s^{-1}, R = 0.004, d_p = 5\text{nm}$ and $\text{Re} = 100$.

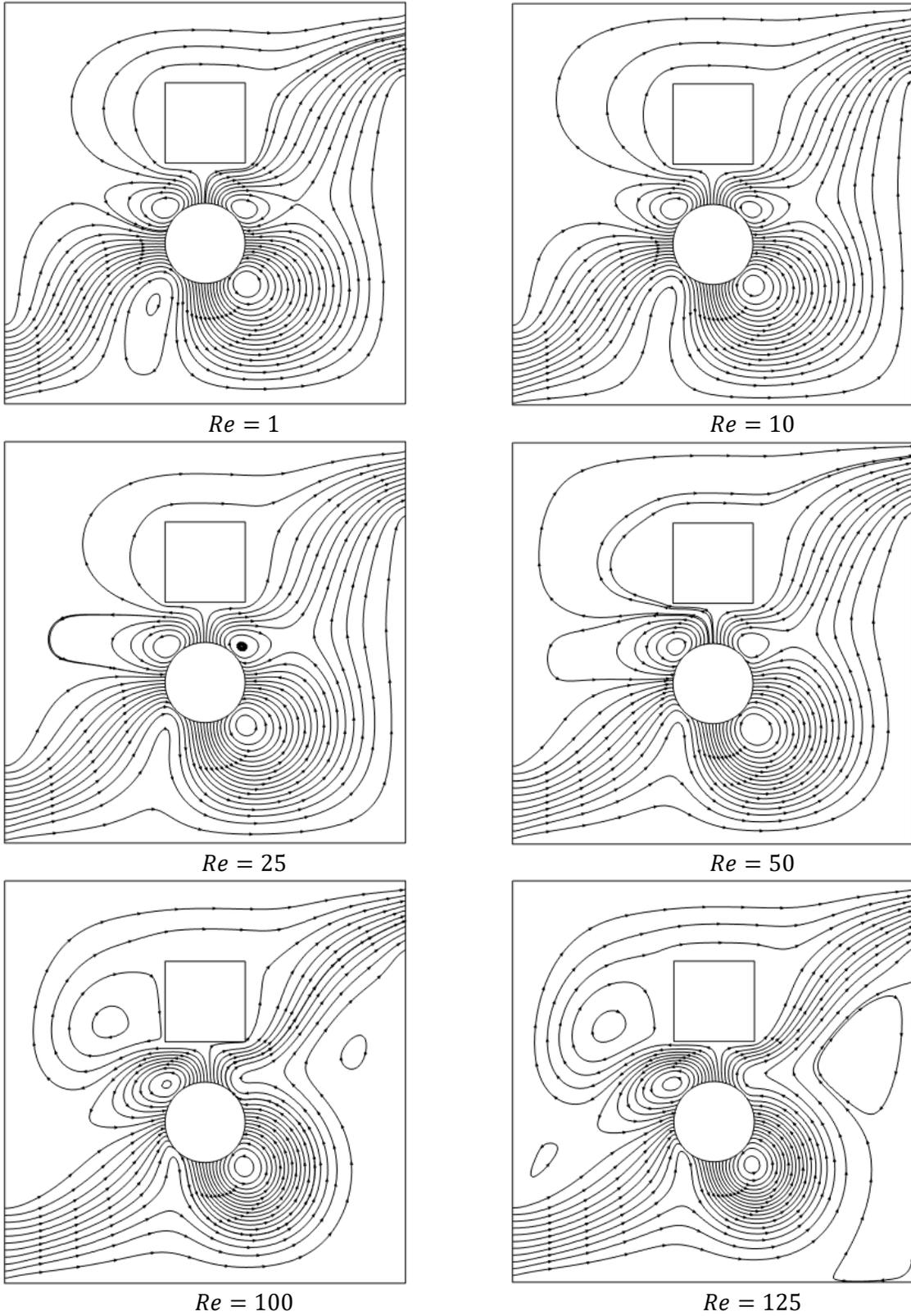
**Fig. 7.** Streamline variation for the variation in Reynolds number (Re) at $\phi = 0.06, \omega = 20s^{-1}, R = 0.004, d_p = 5nm$ and $Ri = 7$.

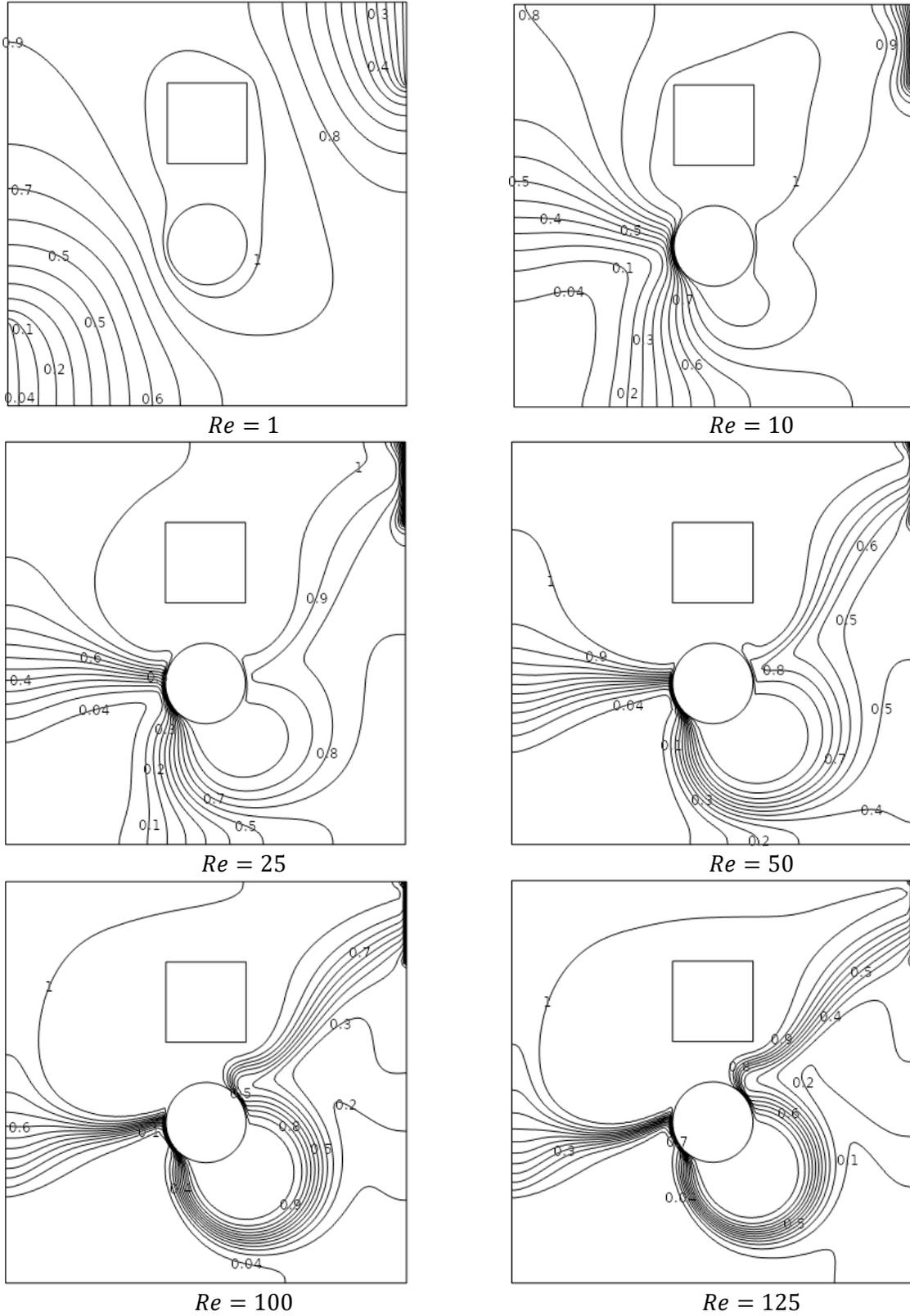

**Fig. 8.** Isothermal contours variation for the variation in Reynolds number (Re) at $\phi = 0.06, \omega = 20 \text{s}^{-1}, R = 0.004, d_p = 5\text{nm}$ and $Ri = 7$.

### Effects of the rotational speed of the hot cylinder ($\omega$) on mixed convection

Fig. 11 and Fig. 12 demonstrates the streamlines and isotherms respectively for different rotational speed of the hot cylinder at $\phi = 0.06, R = 0.004, d_p = 5nm, Re = 100,$ and $Ri = 7$. At $\omega = 1$ rotation per second, the streamlines are very intense near the downside of the hot rotating cylindrical obstacle indicating a very high flow velocity in these regions. Isotherm lines are also very intense around the hot cylindrical obstacle for $\omega = 1$ rotational speed per second, confirming a higher temperature gradient as well as higher heat transfer. At $\omega = 5$ rotation per second, streamlines become denser around the cylinder and generate vortices. With the increased value of the rotational speed, the streamlines become more intense, and generated vortices become more cell centric, that is they rotate with a higher speed than earlier. At $\omega = 25$ rotation per second, two cell-centered vortices rotate near the hot cylinder and the streamlines become denser means higher velocity and higher pressure. From Fig. 12, we can see that the isotherms scatter less dense around the hot obstacles with the increase in the rotational speed of the hot cylinders implies that the temperature gradient decreases, and thus heat transfer decreases for higher rotational speed.

### Effects of inclination angle ($\gamma$) on mixed convection

Fig. 13 demonstrates the streamlines and isotherms for the fixed inclination angle, $\gamma = 15^0$ at $\phi = 0.06, \omega = 20s^{-1}, R = 0.004,$ and $d_p = 5nm$. In part $(a)$ of Fig. 13, a fixed Reynold number, $Re = 100$ is considered for the investigation of changes in streamlines at different Richardson numbers, and also $Ri = 7$ has been taken for the investigation of Reynold number changes. We see that the streamlines around the hot square obstacles and the hot cylinders become denser with the higher Richardson number but streamlines around the hot cylinder decrease with a higher Reynolds number. This confirms that the higher velocity and pressure of the fluid around both hot obstacles for higher Richardson number and lower velocity and pressure for higher Reynolds number is observed and the scenario of higher velocity above the hot square obstacles happens due to the inclination angle.

In part $(b)$ of Fig. 13, isotherm lines around the hot cylinder expand indicating a lower temperature gradient around the hot cylinder and thus, lower heat transfer for higher Richardson numbers at fixed $Re = 100$. A similar scenario is observed for the variation of Reynolds number at fixed $Ri = 7$.

For $\gamma = 30^0$, part $(a)$ of Fig. 14 displays the variations in streamlines and isothermal contours for different Richardson numbers at fixed $Re = 100$, and the part $(b)$ of Fig. 14 displays the variations in streamlines and isothermal contours for different Reynolds numbers at fixed $Ri = 7$. Part $(a)$ of Fig. 14 concludes higher velocity and pressure, with no change in the heat transfer for the variations in the Richardson numbers. Part $(b)$ of Fig. 14 confirms lower velocity and pressure, with increased heat transfer for the variations in the Reynolds number.

Due to the higher inclination angles of $45^0$, and $60^0$, the same scenario can be concluded for the streamlines and isotherms as described above for $\gamma = 15^0$, and $\gamma = 30^0$. Fig. 15 and Fig. 16 show the streamline and isothermal contour variations for different $Ri$ and $Re$ at $\gamma = 45^0$, and $\gamma = 60^0$ respectively.

### Effects of different Re and Ri on the local Nusselt number

Fig. 17 displays the effects of $Re$ and $Ri$ on the local Nusselt number for different inclination angles calculated on the hot wall walls of the obstacles at $\phi = 0.06, R = 0.004, \omega = 20s^{-1},$ and $d_p = 5nm$. Fig. 17($a$) shows the effects of various Reynolds numbers on the Nusselt number at fixed $Ri = 0.1$. The highest Nusselt number is around 27 and is obtained at $Re = 100$ for the inclination angle, $\gamma = 0^0$. In terms of higher to lower local Nusselt number calculated at the hot walls, the inclination angle ($\gamma$) can be serialized as $60^0 < 45^0 < 30^0 < 15^0$. From Fig. 17($b$), we can see that the highest local Nusselt number is calculated for $\gamma = 15^0, Re = 125$ and $Ri = 1$. Fig. 17($c$) shows the effects of Reynold numbers on the local Nusselt number for $Ri = 10$. For this, the scene returns to the same variations for $Ri = 0.1$. Fig. 17(d) displays the effects of different Richardson numbers on the Nusselt number calculated at the hot obstacle walls at $Re = 10$. A very slight decrease in the local Nusselt number is observed for the variations in the Richardson number and thus a linear horizontal line is obtained. The slight decrease may be observed for the changes in the boundary layer thickness near the heated surface with the increasing thermal resistance of the boundary layer. With the increased values of the inclination angles, the value of the local Nusselt number decreases from 11.2 to 2.5 as displayed in the Fig. 17(d). The same scenario is observed for changes in the fixed Reynolds number except for higher $Re$, a higher local Nusselt number is obtained, and vice-versa.

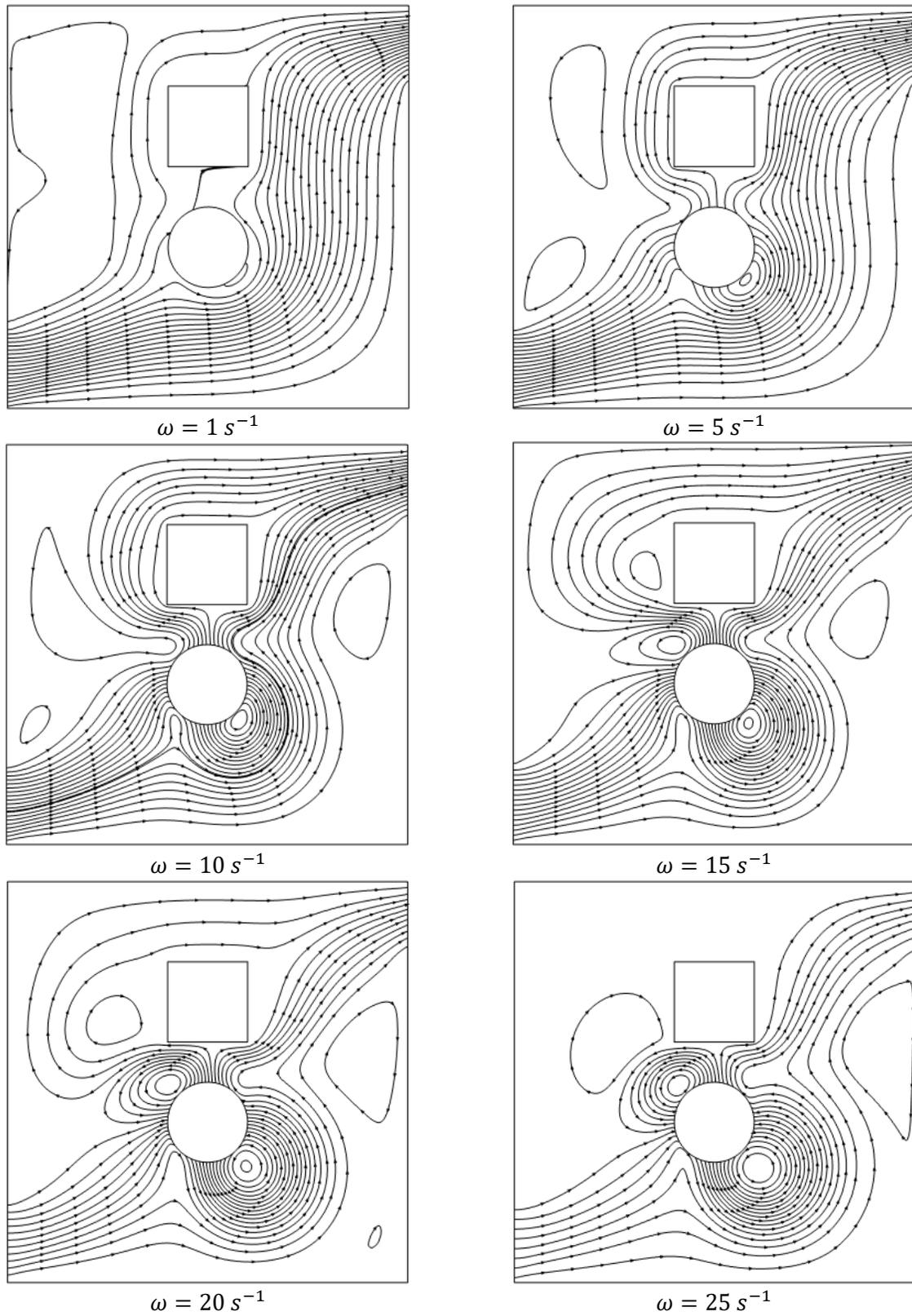

**Fig. 11.** Streamline variation for the variation of the rotational speed (ω) of the cylindrical obstacle (heated) at $\phi = 0.06$, $R = 0.004$, $d_p = 5$nm, $Re = 100$, and $Ri = 7$.

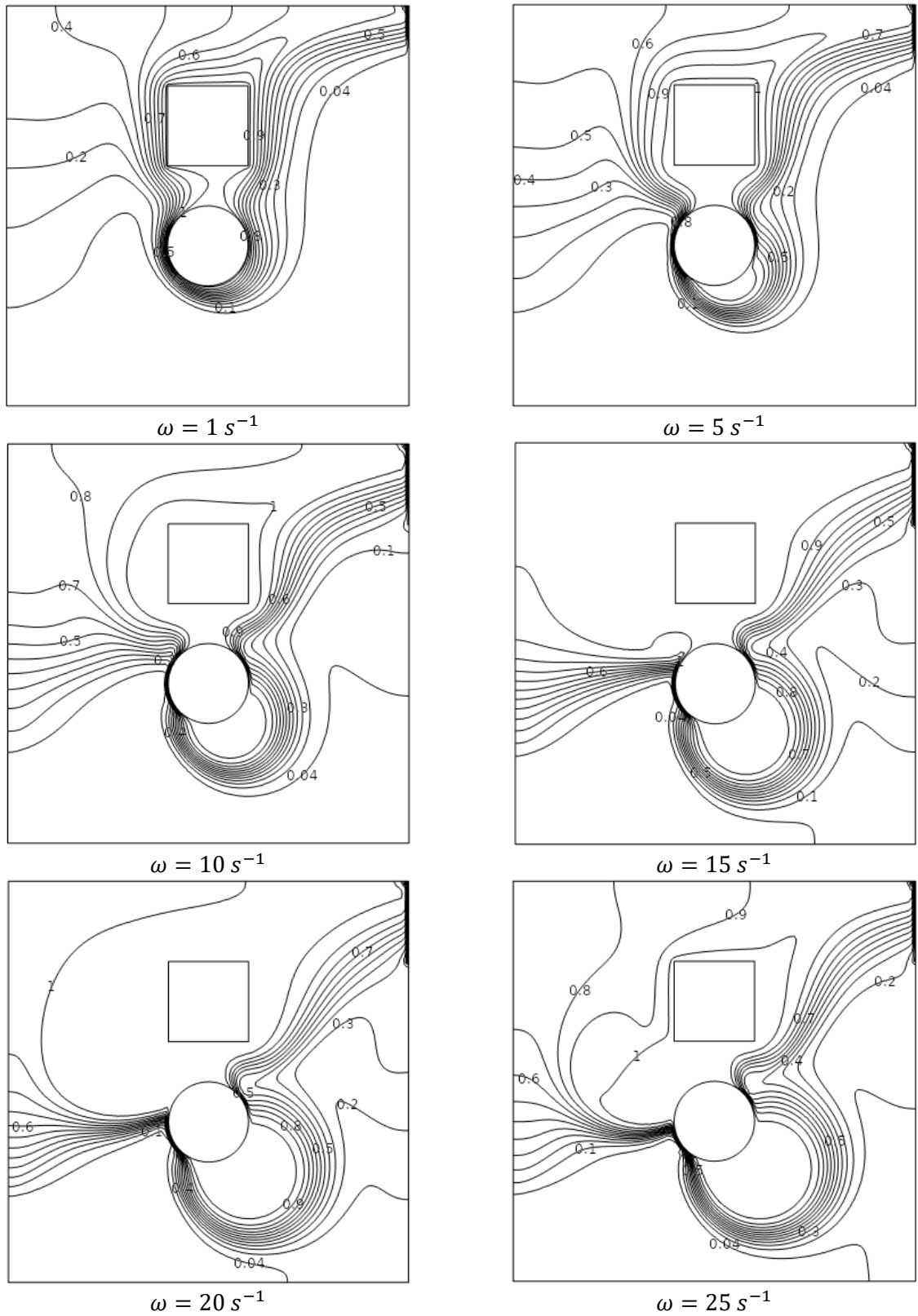

**Fig. 12.** Isothermal contours variation for the variation of the rotational speed (ω) of the cylindrical obstacle (heated) at ϕ = 0.06, R = 0.004, $d_p$ = 5nm, Re = 100, and Ri = 7.

The flow is improved across heated surfaces, and temperature gradients are stronger close to the blocks, which results in increased heat transfer. A larger cavity inclination angle aligns buoyancy and entry fluid forces, producing higher Nusselt numbers, particularly for spinning cylinders. Due to thermal behavior and a significant fluid effect along the wall from vortices, Nusselt number fluctuations with cavity inclination angles are uneven for the square obstruction. Across all Reynolds numbers, Richardson numbers, and inclination angles in the arrangement, the spinning hot cylinder near to the fluid inlet with high velocity and mass flow rate has a considerable influence on Nusselt numbers.

### Effects of changes in nanofluid on the mixed convection

Fig. 18 displays the effects of changes in the nanofluid in the square cavity with the same hot obstacles and same settings. This investigation of changed fluid is done for five different nanofluids including $Al_2O_3$ −water nanofluid with different characteristic properties as mentioned in Table 3. The experiment is done at $\phi = 0.06, \omega = 20s^{-1}, R = 0.004, d_p = 5nm, Re = 100$, and $Ri = 7$. The part $(a)$ of Fig. 18 displays the streamline and isothermal contour variations for $Ag$ −water nanofluid and it shows an intense behavior of the streamlines indicates higher flow velocity and higher pressure in that region with some vortices. Intensive isotherm lines around the hot cylinder obstacle indicate higher heat transfer in the fluids.

The part $(b)$ of Fig. 18 displays the streamline and isothermal contour variations for $Cu$ −water nanofluid and on a comparative view with the $Ag$ −water nanofluid, streamlines are more expanded indicates lower flow but also lower flow velocity and lower pressure near the boundary lines of the square cavity. In the case of heat transfer, it is almost like the $Ag$ −water nanofluid. The part $(c)$ of Fig. 18 displays the streamline and isothermal contour variations for $CuO$ −water nanofluid and the streamline variation and isothermal contours remain the same as for $Cu$ −water nanofluid with the same boundary conditions and same settings.

Effects of $TiO_2$ −water nanofluid in the same square cavity with the same boundary conditions is displayed in the part $(d)$ of Fig. 18. In this case, the streamline intensity is more than the previous three experimented nanofluids indicating higher velocity and pressure with higher heat transfer. Among the five nanofluid cases displayed in Fig. 18, we get higher heat transfer and more diverse streamline variations for $Al_2O_3$ −water nanofluid than others as shown in the part $(e)$ of Fig. 18.

The effective viscosity of the nanofluids considered in this case is computed by the following equation (including $Al_2O_3 -$ water nanofluid):

$$\mu_{nf} = \mu_f(1 + 2.5\phi) \tag{24}$$

Nanofluids have various distinct characteristics that set them apart from dispersions of $mm$ or $\mu m$ sized particles.

Table 3.
Physical properties of different nanoparticles.

| Properties | $H_2O$ | $Ag$ | $Cu$ | $CuO$ | $TiO_2$ | $Al_2O_3$ |
|---|---|---|---|---|---|---|
| $C_p$ | 4179 | 235 | 385 | 531.8 | 686.2 | 765 |
| $\rho$ | 997.1 | 10500 | 8933 | 6320 | 4250 | 3970 |
| $k$ | 0.613 | 429 | 401 | 76.5 | 8.95 | 25 |
| $\beta (.10^{-5})$ | 21 | 1.89 | 1.67 | 1.8 | 0.9 | 0.85 |

Water is the starting point in every scenario.

Nanofluids have been proven to have greater thermal conductivities than traditional cooling liquids such as water, kerosene, ethylene glycol, and microfluidics. Furthermore, nanofluids do not obstruct flow channels and only cause a little pressure to drop during flow, making them ideal for heat transfer applications.

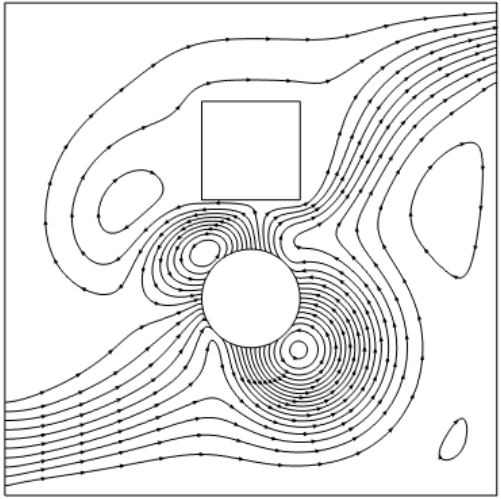 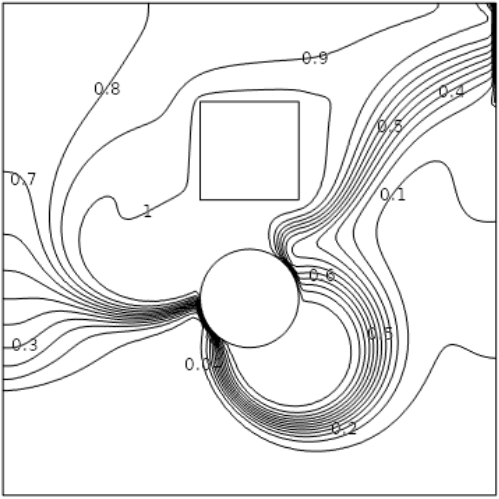

(**a**) Streamline and isotherm contours variation for Ag-water nanofluid at $\phi = 0.06, \omega = 20s^{-1}, R = 0.004$.

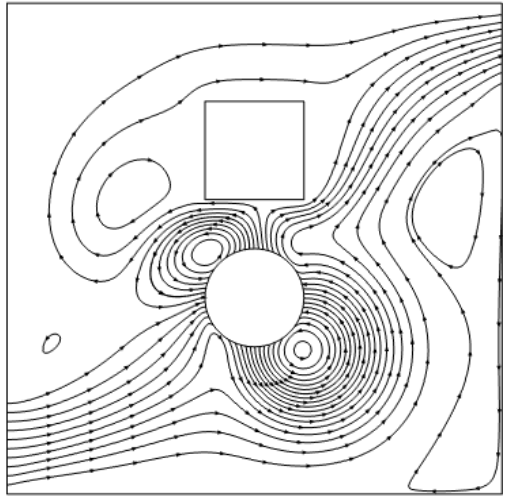 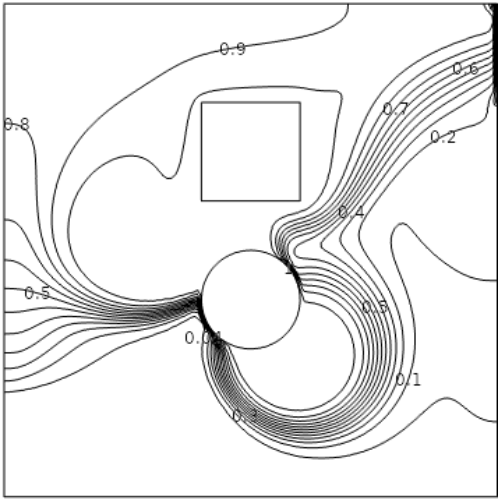

(**b**) Streamline and isotherm contours variation for Cu-water nanofluid at $\phi = 0.06, \omega = 20s^{-1}, R = 0.004$.

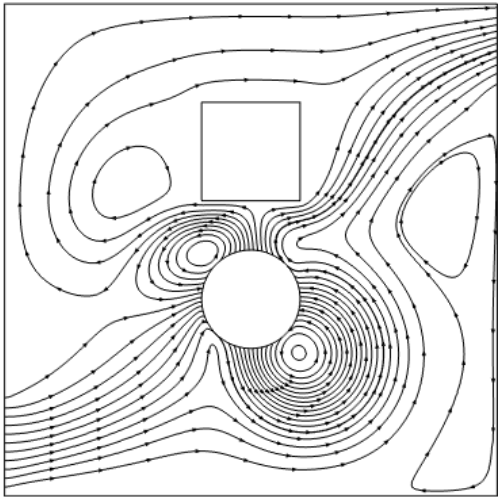 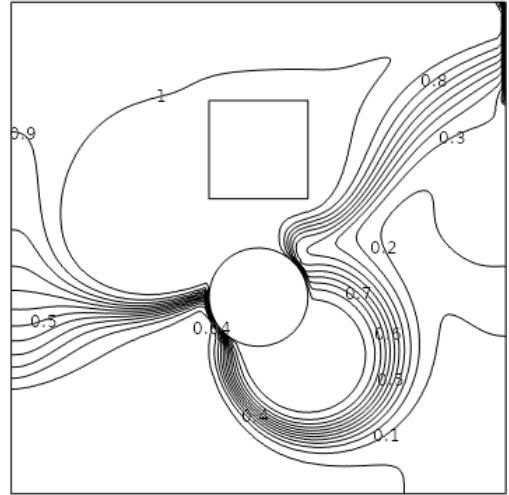

(**c**) Streamline and isotherm contours variation for CuO-water nanofluid at $\phi = 0.06, \omega = 20s^{-1}, R = 0.004$.

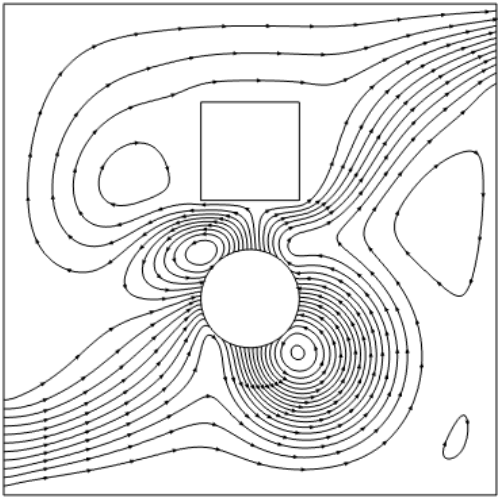 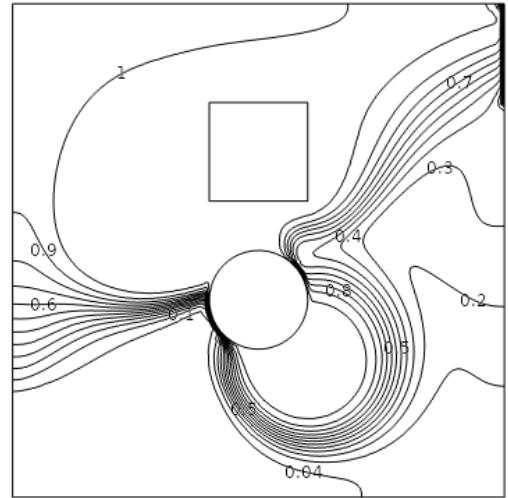

(**d**) Streamline and isotherm contours variation for $TiO_2$-water nanofluid at $\phi = 0.06, \omega = 20s^{-1}, R = 0.004$.

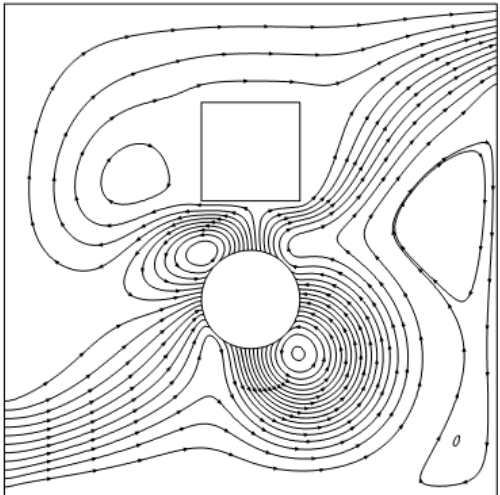 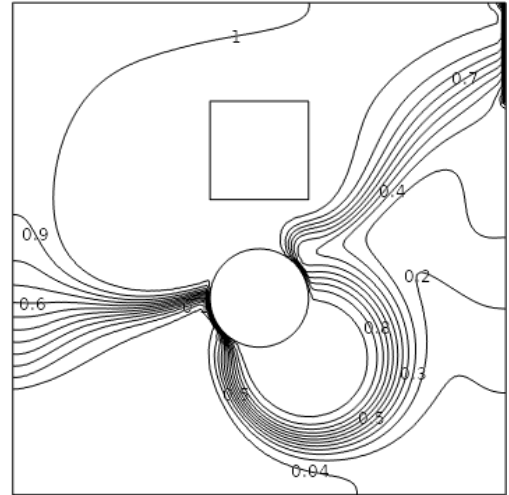

(**e**) Streamline and isotherm contours variation for $Al_2O_3$-water nanofluid at $\phi = 0.06, \omega = 20s^{-1}, R = 0.004$.

**Fig. 18.** Streamline and isotherm contours variation for different nanofluids at Re = 100 and Ri = 7 ($25^0C, d_p = 5nm$).

**Effects of wavy walls of the square cavity on the mixed convection**

The effect of wavy walls in the square is displayed in the following Fig. 19 and the boundary conditions of the wavy walls, in this case, are applied using the COMSOL Multiphysics software package. Six different cases of wavy wall situations have been considered to calculate the streamline and isothermal variations for the $Al_2O_3$ −water nanofluid in the square cavity $at\ \phi = 0.06, \omega = 20s^{-1}, R = 0.004, d_p = 5nm, Re = 100,\ and\ Ri = 7$.

For the down wavy wall, the streamline becomes intensively denser and uniformly scattered on the downside of the rotating cylinders with two vortices. Isothermal contour lines are intense near the middle and upper regions of the hot square and cylindrical obstacles. This confirms the higher flow velocity and higher pressure in the downside of the square cavity and higher heat transfer for the hot square obstacle. For the top wavy wall, the streamlines become intensively denser and uniformly scattered on the right side of the rotating cylinders with one vortex. This indicates the higher flow velocity and higher pressure in the right side of the square cavity and higher heat transfer for the hot cylindrical obstacle, as the isotherm lines are intensively scattered around the rotating cylinder. For the down and top wavy walls, we get about the same type of variations of the streamlines and isotherms as we already explained for the down wavy wall. From Fig. 19, we can see that the streamlines and isothermal contours are almost similar for the left wavy wall, left and right wavy walls, and all wavy walls cases. The streamline and isothermal variations are more intense for all wavy walls than in the other two cases, indicating higher velocity and pressure with higher heat transfer.

**Effects of obstacles position on the mixed convection**

Fig. 20 and Fig. 21 display the streamlines and isotherms respectively for different positions of the hot square and cylindrical obstacles $at\ \phi = 0.06, \omega = 20s^{-1}, R = 0.004, d_p = 5nm, Re = 100,\ and\ Ri = 7$. The part $(a)$ of Fig. 20 and Fig. 21 show the effects of changes in the horizontal position of the heated obstacles in the streamlines and isotherms respectively. Similarly, part $(b)$ of Fig. 20 and Fig. 21 show the effects of changes in the vertical position of the heated obstacles in the streamlines and isotherms respectively.

For the right-down position of the hot rotating cylinders, the streamlines are intensively scattered around both heated obstacles with two vortices in the middle of the square cavity and this results in higher velocity and higher pressure around the whole square cavity. The intensive isotherm lines near the inlet and the heated rotating cylinder indicate a higher temperature gradient and thus, higher heat transfer in that region. The middle-down position of the heated rotating cylinder confirms the normal situation of fluid flow as described earlier. And the left-down position of the heated cylinder shows the intensive streamline variations in the lower and right parts of the square cavity. Moreover, two vortices are generated on the right side of the cavity. Isothermal contour variations are more intensive around the heated rotating cylinders indicating more heat transfer than right-down and middle-down positions.

The converse change of position of the rotating cylinders displays different velocity streamlines and isotherms than the down position of the heated rotating cylinders. In the left-top position, the streamlines are uniformly scattered around the heated cylinder with two vortices ensuring a higher velocity and pressure in the top region of the square cavity. And the isothermal contour lines are dispersed intensively in the lower position of the cavity. The middle-top position of the rotating cylinder displays almost the same scenario as the left-top position but for the right-top position of the rotating cylinders, we see some flow velocities in the down area of the heated square obstacle.

Part (*b*) of Fig. 21 shows the changes in the streamlines and isotherms for the horizontal position of the heated obstacles and variations for six different vertical positions. In the right-top position of the heated rotating obstacles, the streamlines are intense around the heated square and the isotherm lines are denser in the region lower middle of the cavity. The right-middle position ensures more velocity and pressure and more heat transfer than the right-top and right-down position. On the horizontal inverse position of the heated cylindrical obstacle, the left-down position ensures better flow around the cavity with three vortices and higher heat transfer from the rotating cylinder.

Fig. 22 demonstrates the streamline and isothermal contour variation for different shapes of the obstacles inside the square cavity with the same settings at $\phi = 0.06, R = 0.004, \omega = 20s^{-1}, d_p = 5nm, Re = 100,$ *and* $Ri = 7$. The six different shapes of the obstacles considered in this investigation are square-square, square-triangle, triangle-triangle, triangle-circle, circle-circle, and rectangle-trapezoid. A further detailed study of these geometries is the planned expansion of this work.

## Conclusion and future directions

The stationary heated square form impediment is responsible for natural convection, while the hot revolving cylinder is responsible for forced convection, resulting in a mixed convection problem. Variations in the Richardson number, Reynolds number, cylinder rotational speed, inclination angle, the position of obstacles, changes in the nanofluid, and effects of wavy walls on the fluid flow and thermal fields in the enclosure, as well as changes in the Nusselt number, were investigated in several cases. The following items are derived from the numerical simulation's study of streamlines, temperature lines, and Nusselt diagrams:

- Increasing the volume fraction of the nanoparticles with smaller diameters leads the ratio of thermal conductivity to rise for both barriers and all ranges of the parameters in this investigation. Consequently, increasing the Nusselt number means increasing heat transmission.
- Increased forced convection produces a slight reduction in heat transmission, as well as decreased velocity and pressure inside the enclosure, as the Richarson number rises.
- In all circumstances, when the Reynolds number rises, the Nusselt number rises, resulting in greater heat transfer, as well as increased velocity and pressure around the enclosure.
- As the rotating speed of the heated cylinder increases, intense streamlines with vortices form, increasing velocity and pressure while decreasing heat transmission.
- Increasing the inclination angle of the enclosure causes the Nusselt number computed at the heated barriers to drop in all ranges of parameters in this study, implying that heat transmission is reduced while velocity and pressure surrounding the enclosure are increased.
- We altered the nanofluid and found that $Al_2O_3$ −water nanofluid indicates superior mixed heat transfer than $Ag$ −water, $Cu$ −water, $CuO$ −water, and $TiO_2$ −water nanofluids with the same settings.
- In the instance of wavy walls, we can see that as the length of the wavy walls borders of the enclosure increases, so does the velocity, pressure, and heat transmission.
- The positional shift of the obstacles examined has a substantial impact on the streamlines and isotherm lines. Higher velocity, pressure, and heat transmission inside the enclosure are ensured by the cylinder's right-down vertical and left-down horizontal positions.

For future development on this project, the following recommendation might be made:
- Within the enclosure, magnetic fluid or electrically conducting fluid can be used to conduct the investigation. Also, the investigation can be extended to hybrid Nanomaterials.
- Characterization of nanofluids adjusting different boundary conditions of the enclosure walls may be introduced. For environmental protection, this analysis would be more beneficial.
- For a better understanding of the effect of either heat transfer irreversibility or fluid friction irreversibility on entropy formation, the entropy generation rate may be computed and shown as Bejan lines.

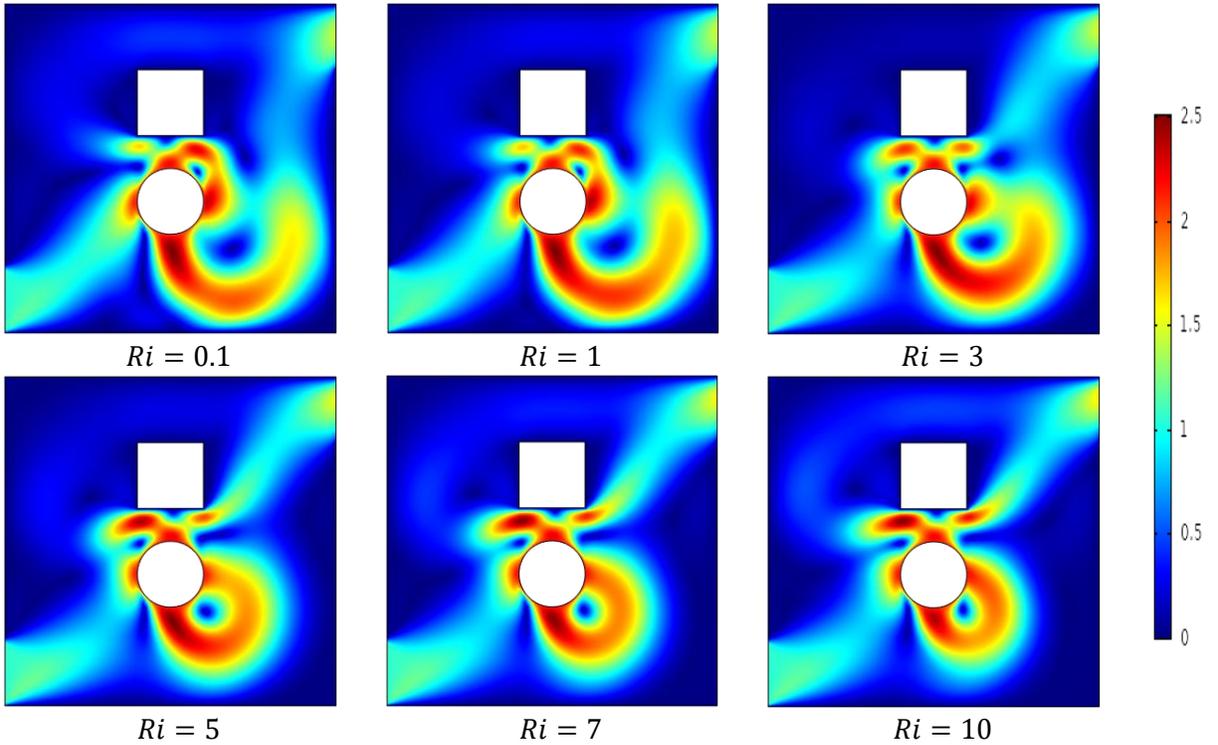

(**a**) Velocity surface at Re = 100.

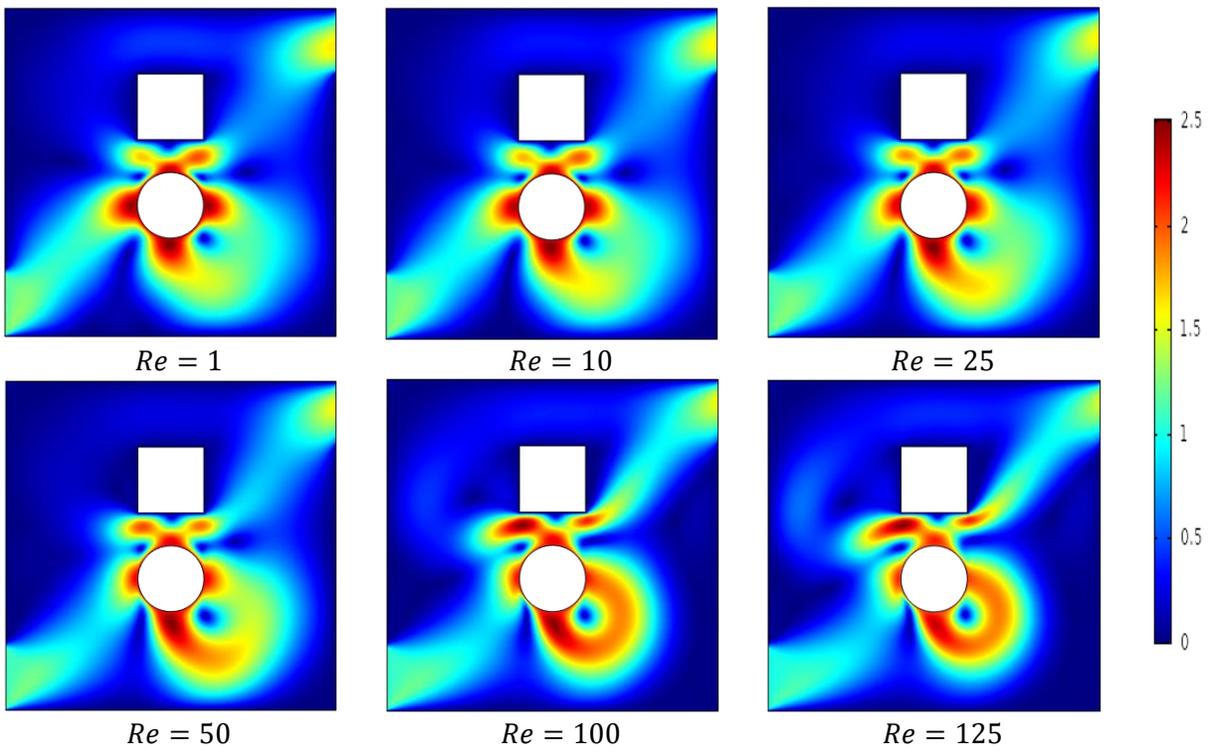

(**b**) Velocity surface at Ri = 7.

**Fig. 9.** Variations of velocity surface for the variation in Reynolds number (Re) and Richardson number (Ri) at $\phi = 0.06, \omega = 20 s^{-1}, R = 0.004$ and $d_p = 5 nm$.

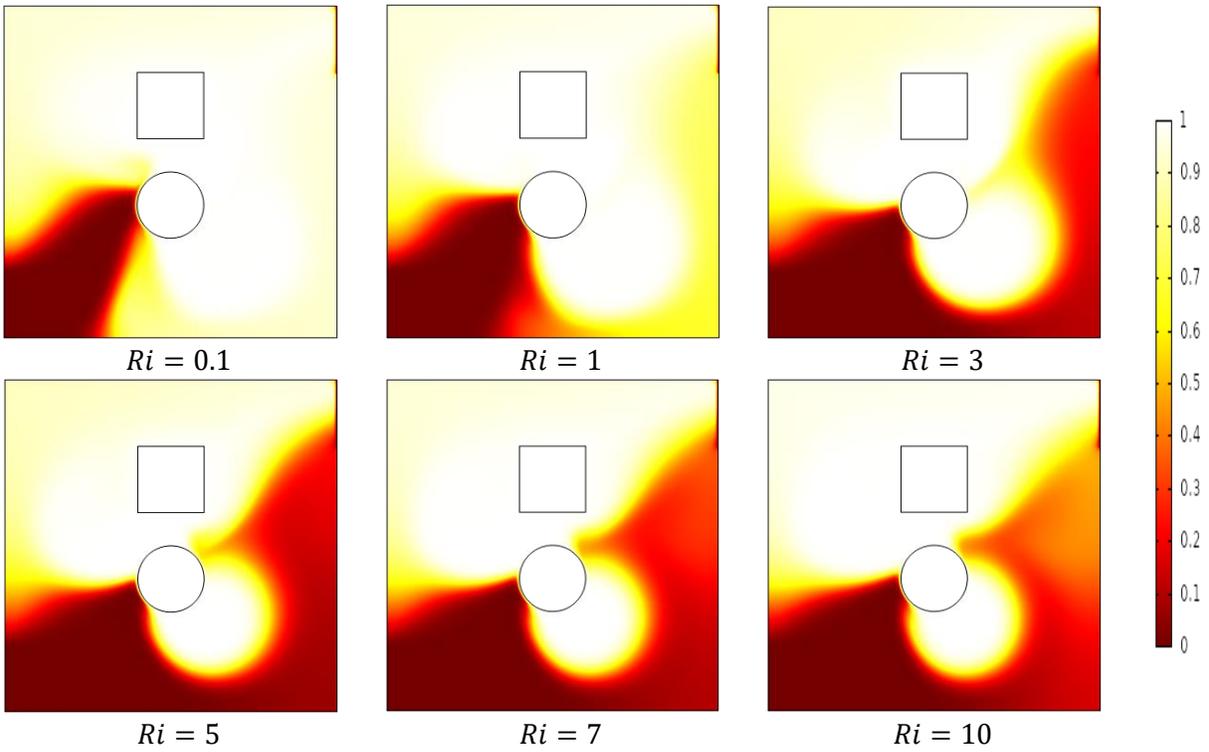
(**a**) Temperature field variation at Re = 100.

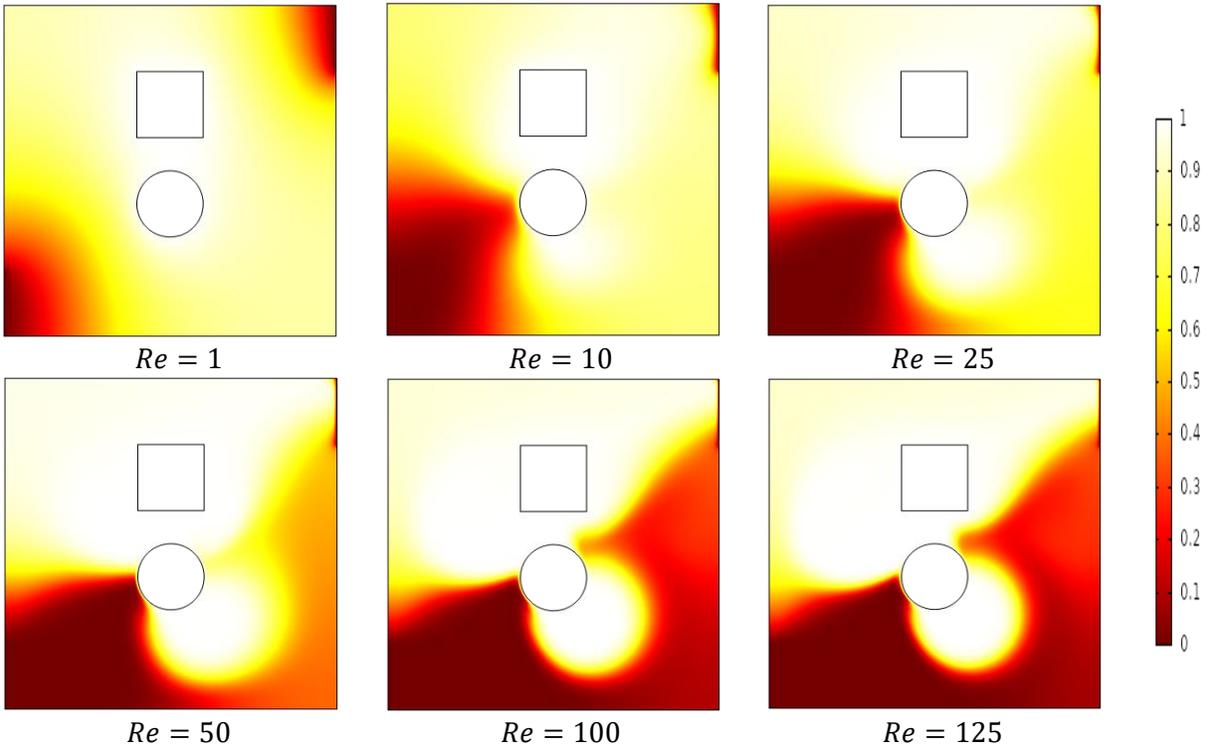
(**b**) Temperature field variation at Ri = 7.

**Fig. 10.** Variations of temperature field for the variation in Reynolds number (Re) and Richardson number (Ri) at $\phi = 0.06, \omega = 20s^{-1}, R = 0.004$ and $d_p = 5nm$.

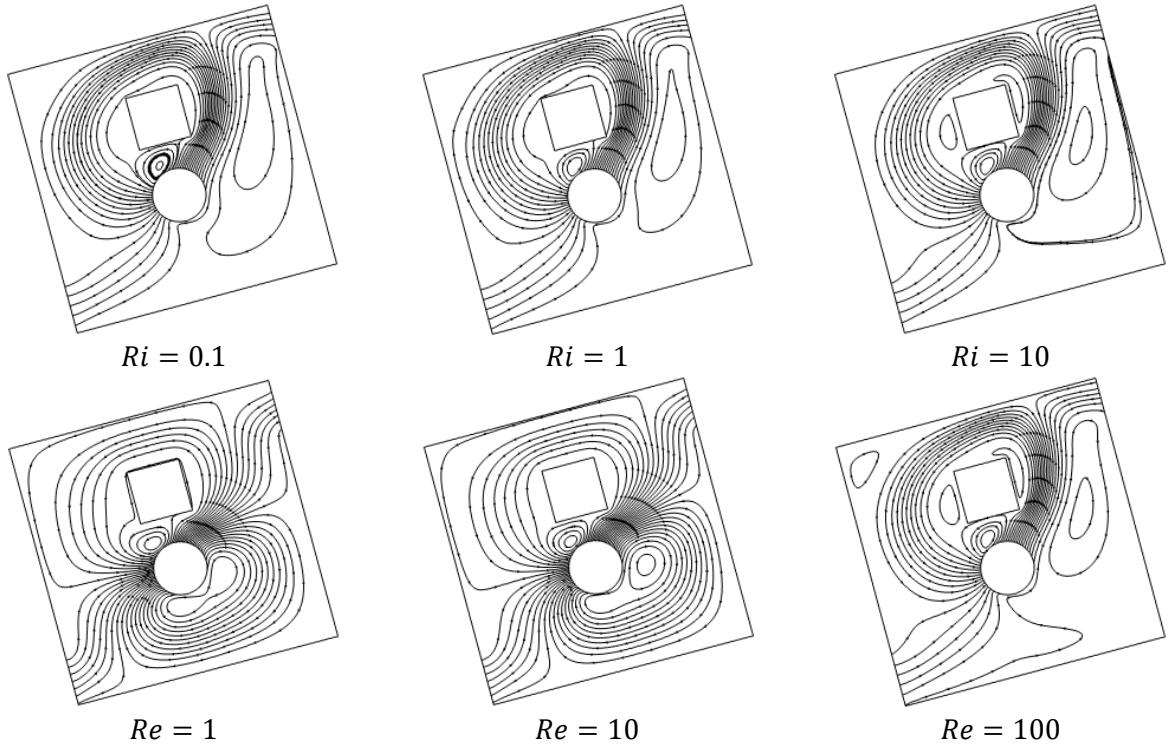

(**a**) Effects of inclination angle on velocity streamlines: (i) First row, Re = 100. (ii) Second row, Ri = 7.

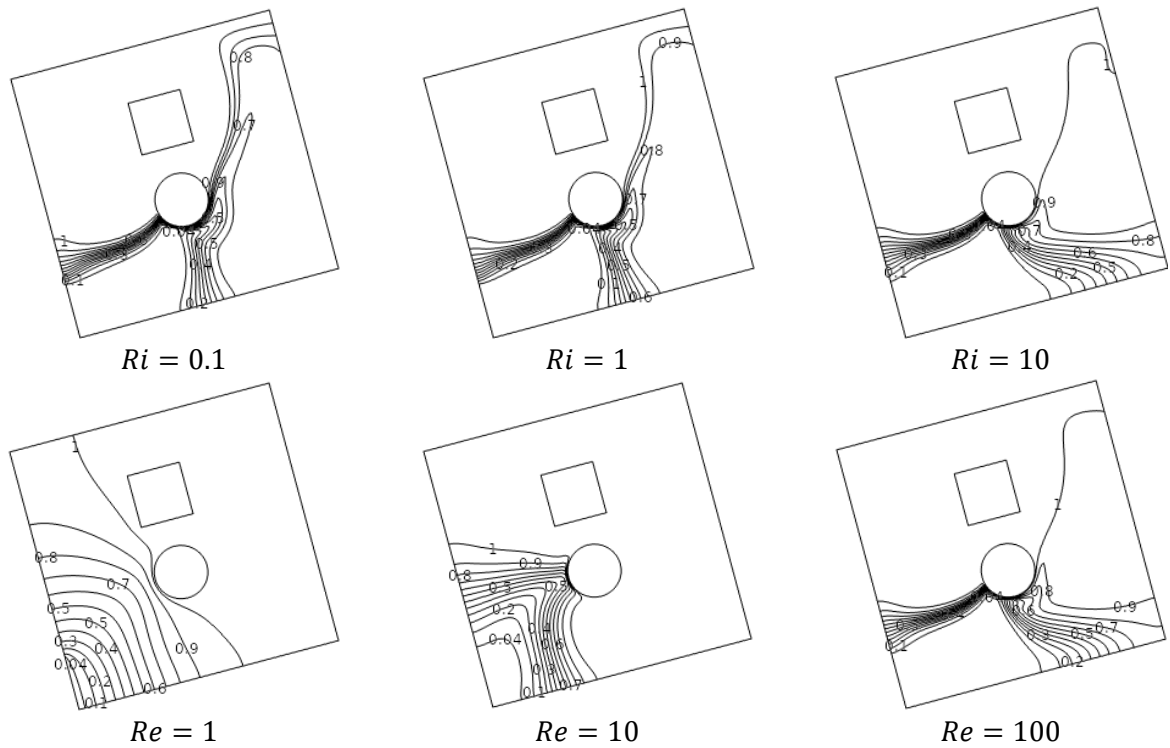

(**b**) Effects of inclination angle on isothermal contours: (i) First row, Re = 100. (ii) Second row, Ri = 7.

**Fig. 13.** Effects of inclination angle $\gamma = 15^0$ on the velocity streamline and isothermal contours at $\phi = 0.06, R = 0.004, \omega = 20s^{-1}$ and $d_p = 5nm$.

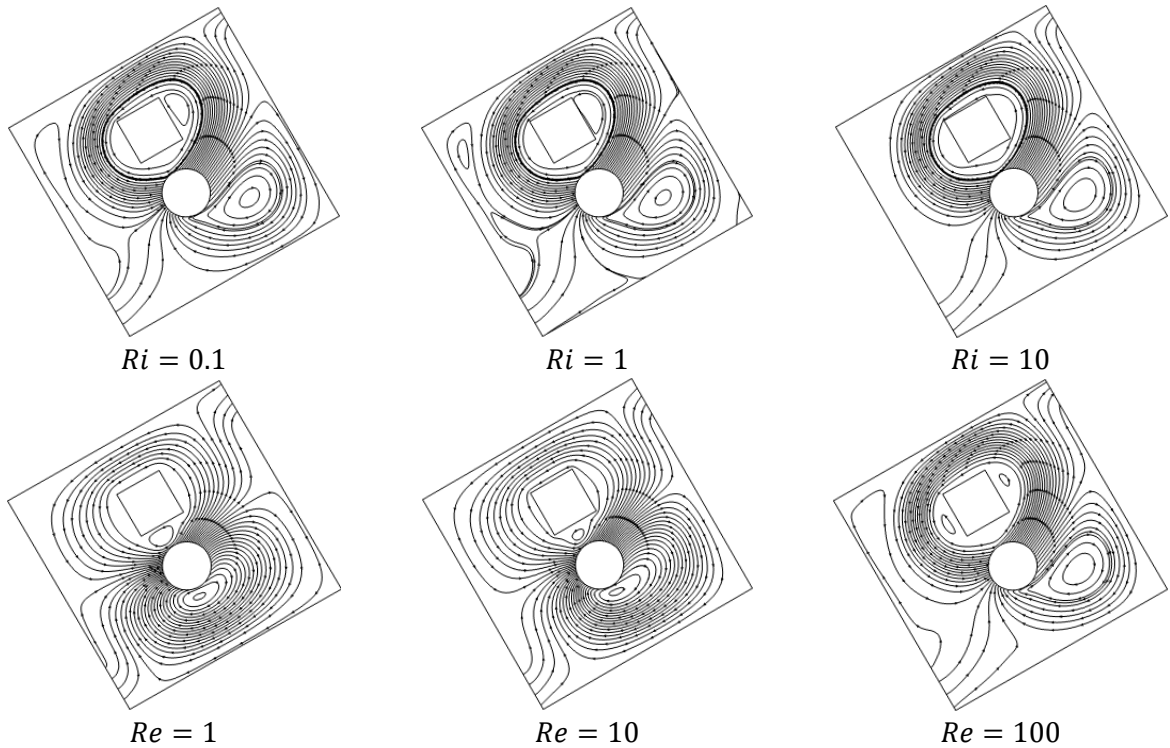

(**a**) Effects of inclination angle on velocity streamlines: (i) First row, Re = 100. (ii) Second row, Ri = 7.

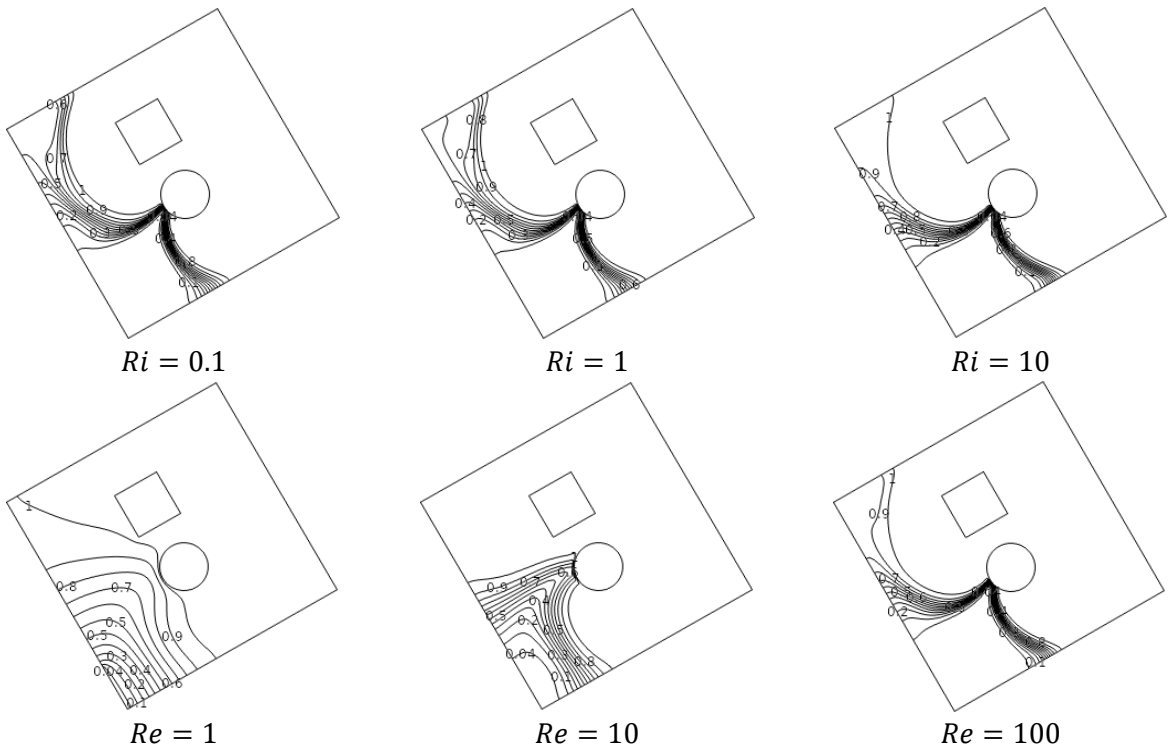

(**b**) Effects of inclination angle on isothermal contours: (i) First row, Re = 100. (ii) Second row, Ri = 7.

**Fig. 14.** Effects of inclination angle $\gamma = 30^0$ on the velocity streamline and isothermal contours at $\phi = 0.06, R = 0.004, \omega = 20s^{-1}$ and $d_p = 5nm$.

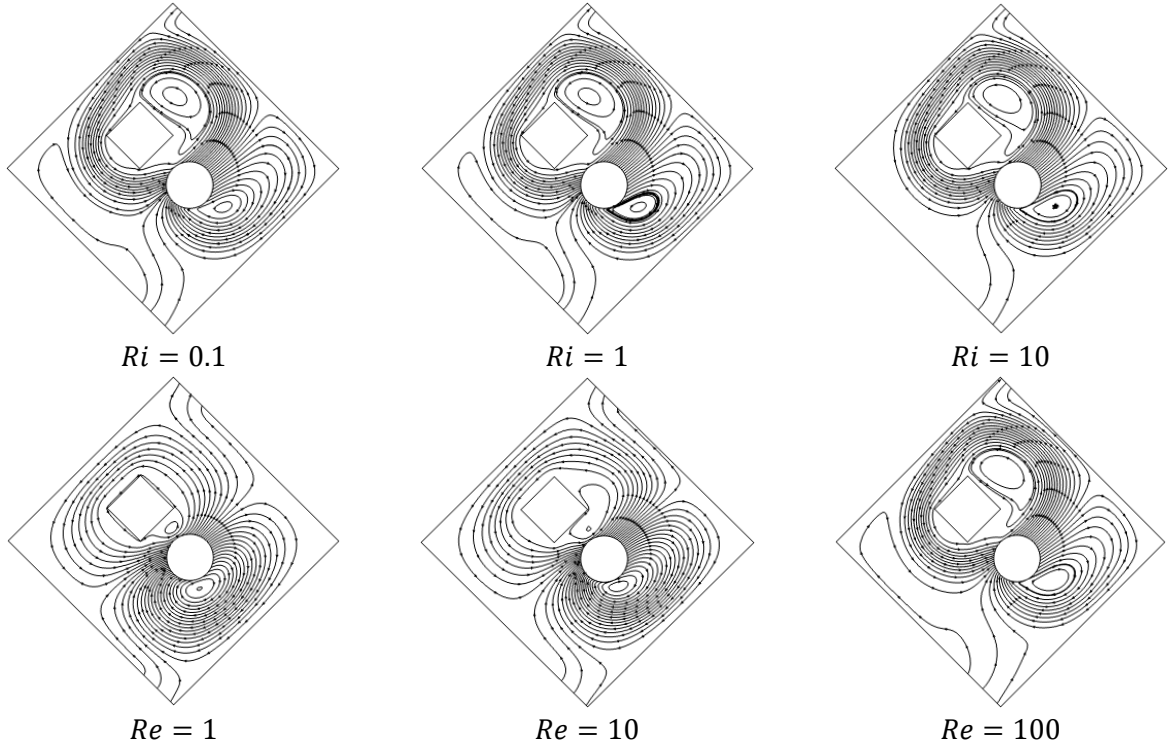

(**a**) Effects of inclination angle on velocity streamlines: (i) First row, Re = 100. (ii) Second row, Ri = 7.

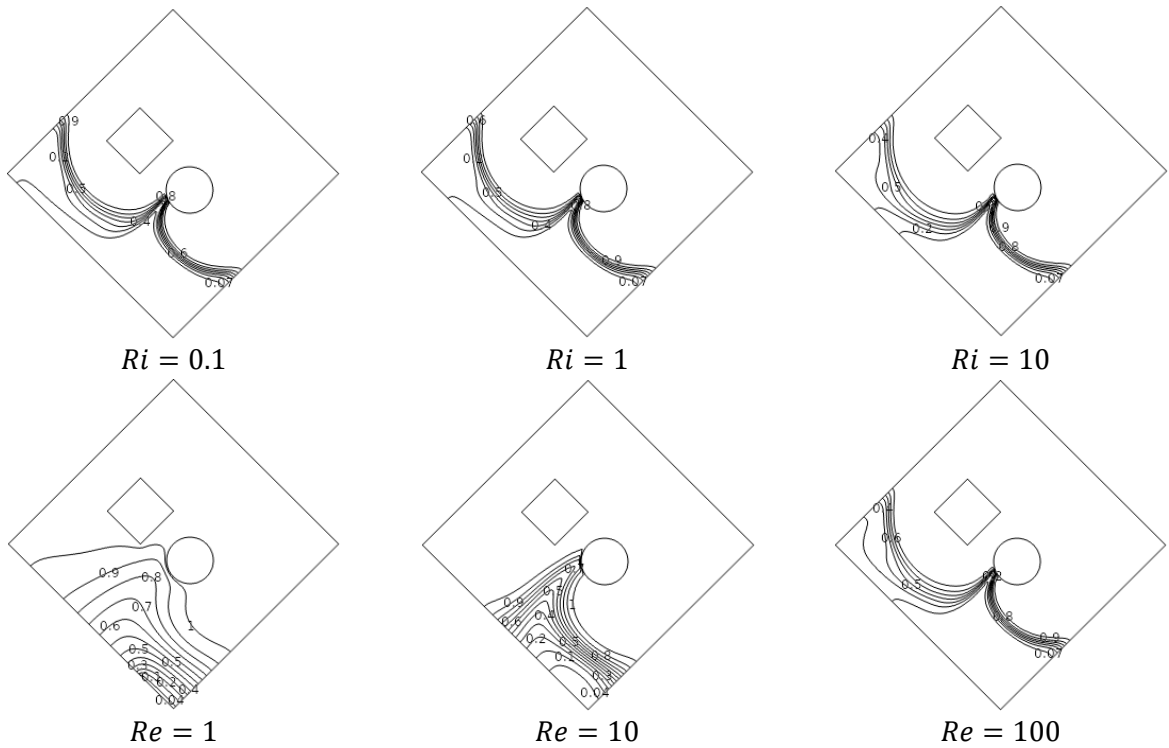

(**b**) Effects of inclination angle on isothermal contours: (i) First row, Re = 100. (ii) Second row, Ri = 7.

**Fig. 15.** Effects of inclination angle $\gamma = 45^0$ on the velocity streamline and isothermal contours at $\phi = 0.06, R = 0.004, \omega = 20 s^{-1}$ and $d_p = 5 nm$.

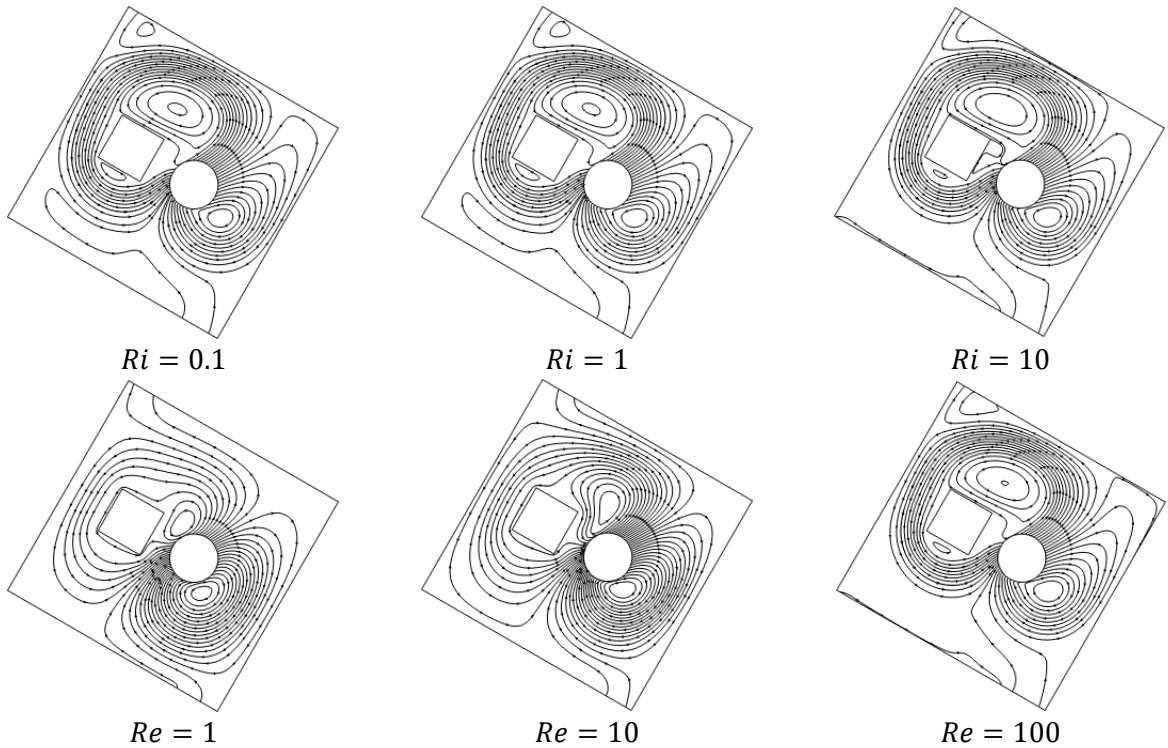

(**a**) Effects of inclination angle on velocity streamlines: (i) First row, Re = 100. (ii) Second row, Ri = 7.

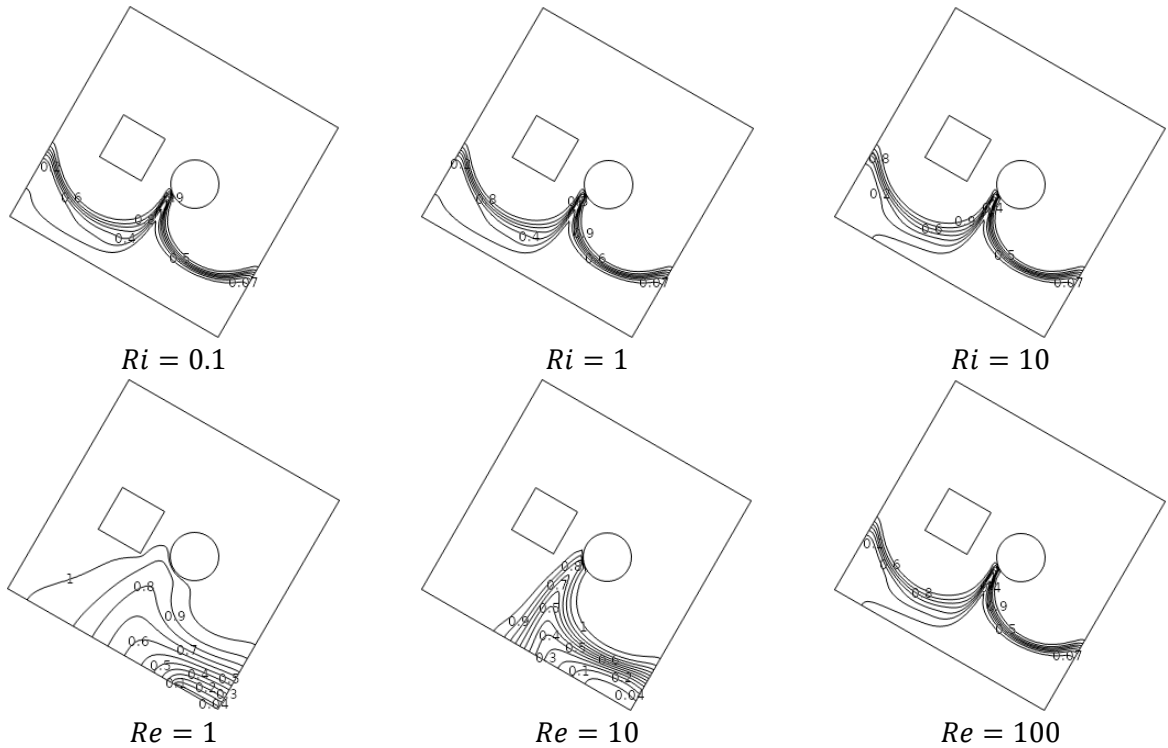

(**b**) Effects of inclination angle on isothermal contours: (i) First row, Re = 100. (ii) Second row, Ri = 7.

**Fig. 16.** Effects of inclination angle $\gamma = 60^0$ on the velocity streamline and isothermal contours at $\phi = 0.06, R = 0.004, \omega = 20s^{-1}$ and $d_p = 5nm$.

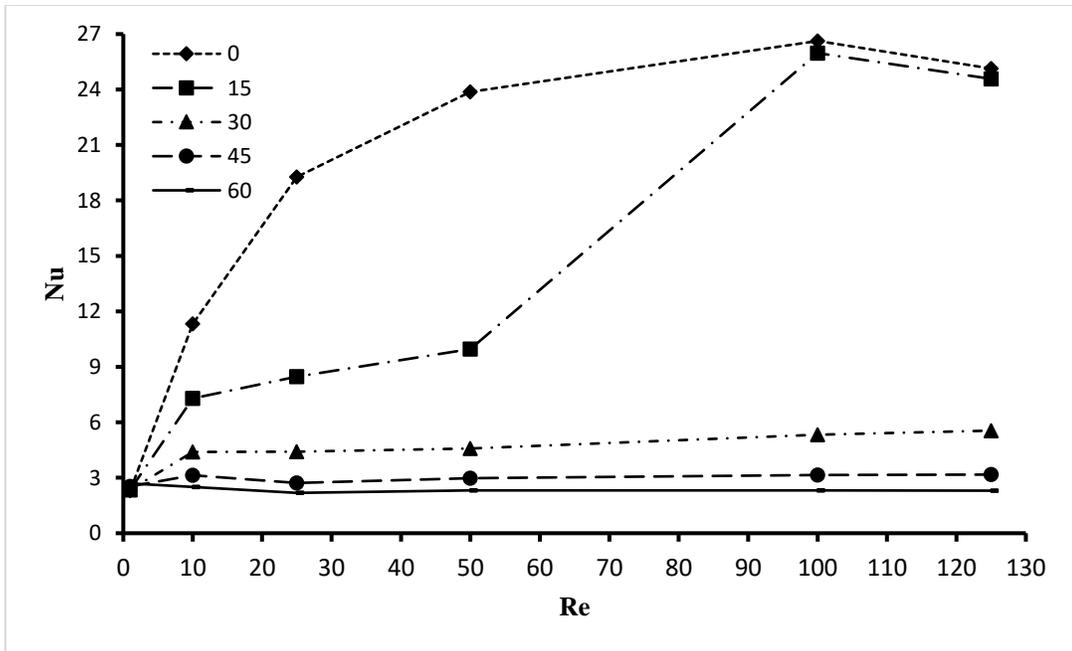

**(a)** Effects of Reynolds number (Re) on the Nusselt number calculated on the hot walls for different inclination angles at $\phi = 0.06, R = 0.004, \omega = 20 s^{-1}, d_p = 5nm$ and $Ri = 0.1$.

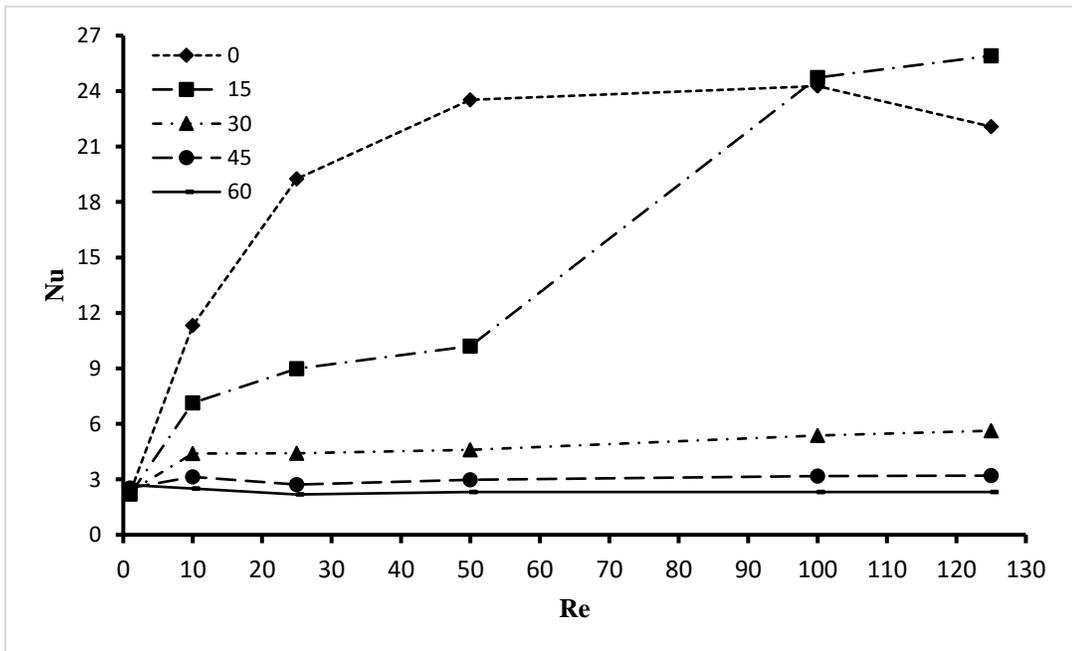

**(b)** Effects of Reynolds number (Re) on the Nusselt number calculated on the hot walls for different inclination angles at $\phi = 0.06, R = 0.004, \omega = 20 s^{-1}, d_p = 5nm$ and $Ri = 1$.

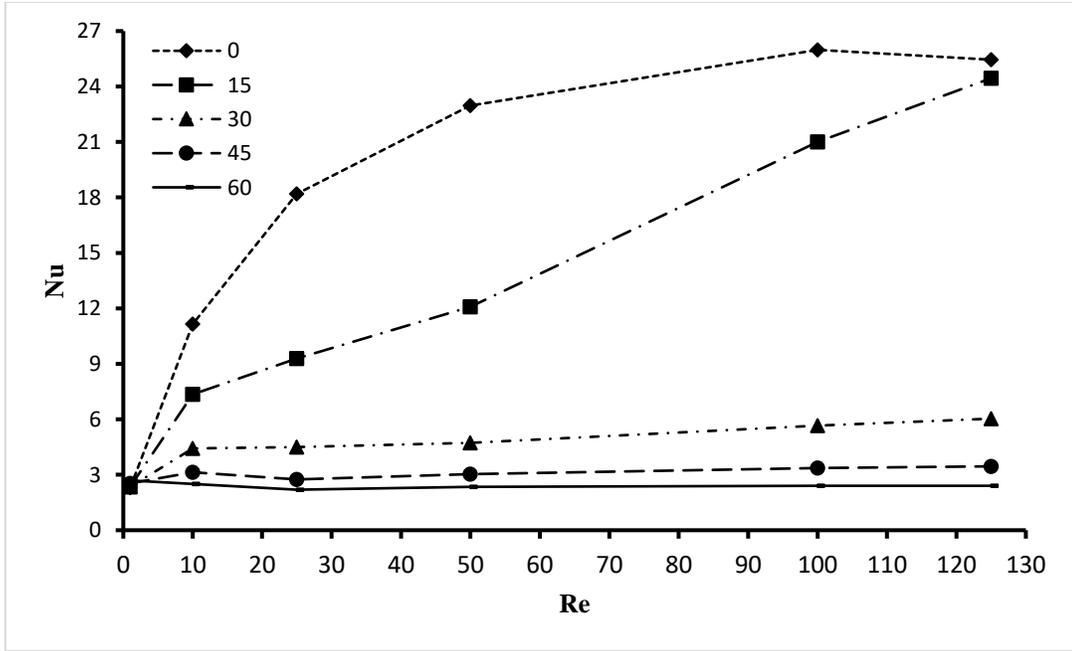

(c) Effects of Reynolds number (Re) on the Nusselt number calculated on the hot walls for different inclination angles at $\phi = 0.06, R = 0.004, \omega = 20s^{-1}, d_p = 5nm$ and $Ri = 10$.

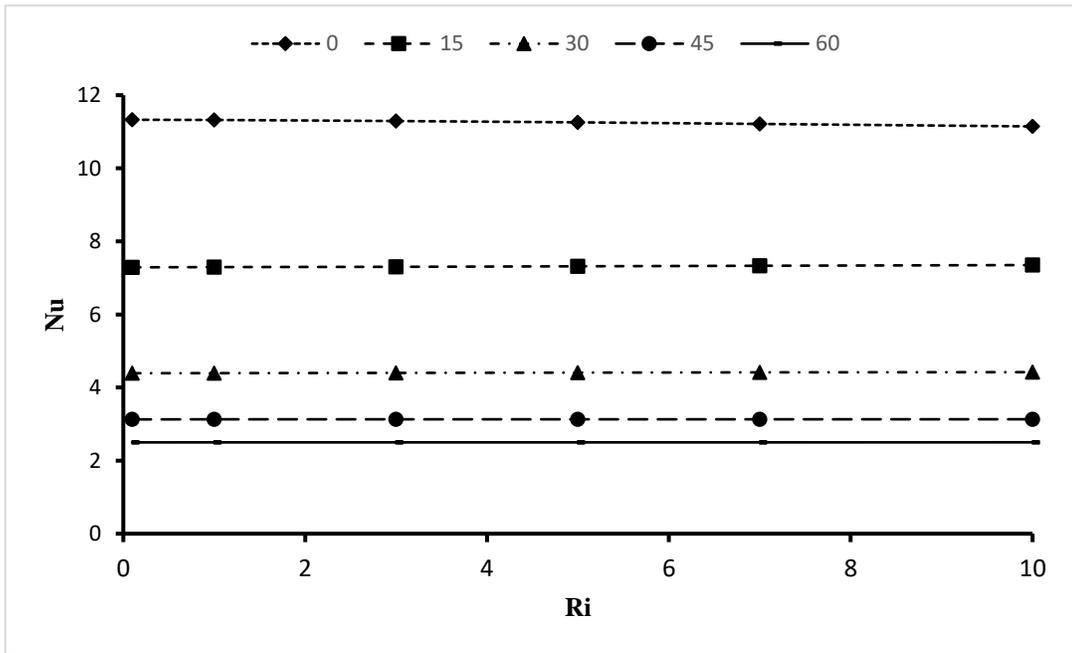

(d) Effects of Richardson number (Ri) on the Nusselt number calculated on the hot walls for different inclination angles at $\phi = 0.06, R = 0.004, \omega = 20s^{-1}, d_p = 5nm$ and $Re = 10$.

**Fig. 17.** Effects of Reynolds number ($Re$) and Richardson number (Ri) on the Nusselt number calculated on the hot walls for different inclination angles at $\phi = 0.06, R = 0.004, \omega = 20s^{-1}, d_p = 5nm$ and $Re = 10$.

(**a**) Effect of wavy walls on velocity streamlines.

(**b**) Effect of wavy walls on isothermal contours.

**Fig. 19.** Effects of wavy walls on the velocity streamline and isothermal contours at $\phi = 0.06, \omega = 20\text{s}^{-1}, R = 0.004, d_p = 5\text{nm}, \text{Re} = 100,$ and $\text{Ri} = 7$.

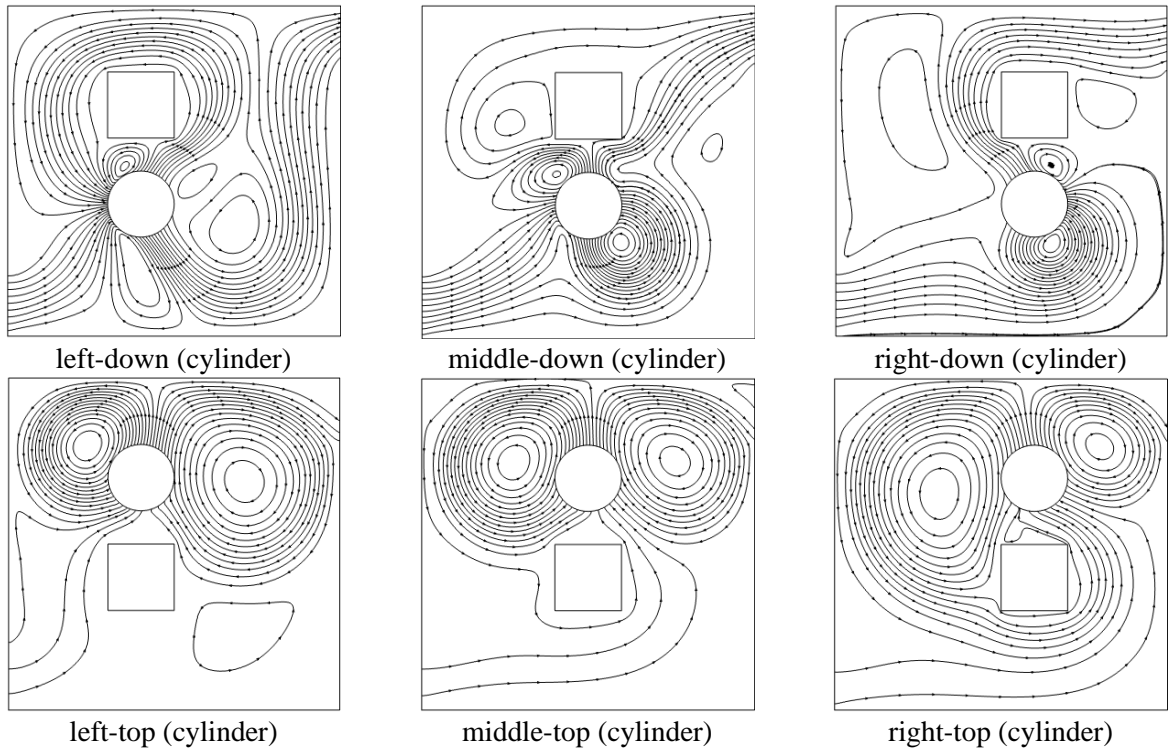
(a) Effect of horizontal position change on velocity streamlines.

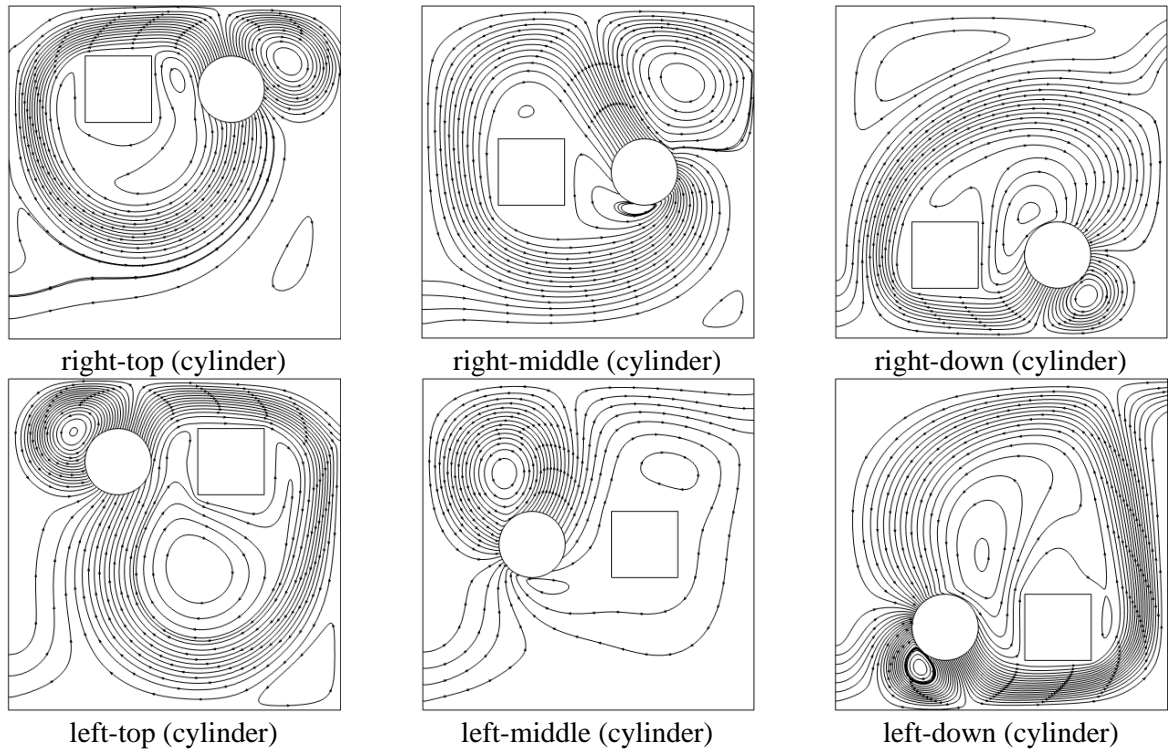
(b) Effect of vertical position change on velocity streamlines.

**Fig. 20.** Variations of velocity streamline for the change of position of the obstacles at $\phi = 0.06, \omega = 20s^{-1}, R = 0.004, d_p = 5nm, Re = 100,$ and $Ri = 7$.

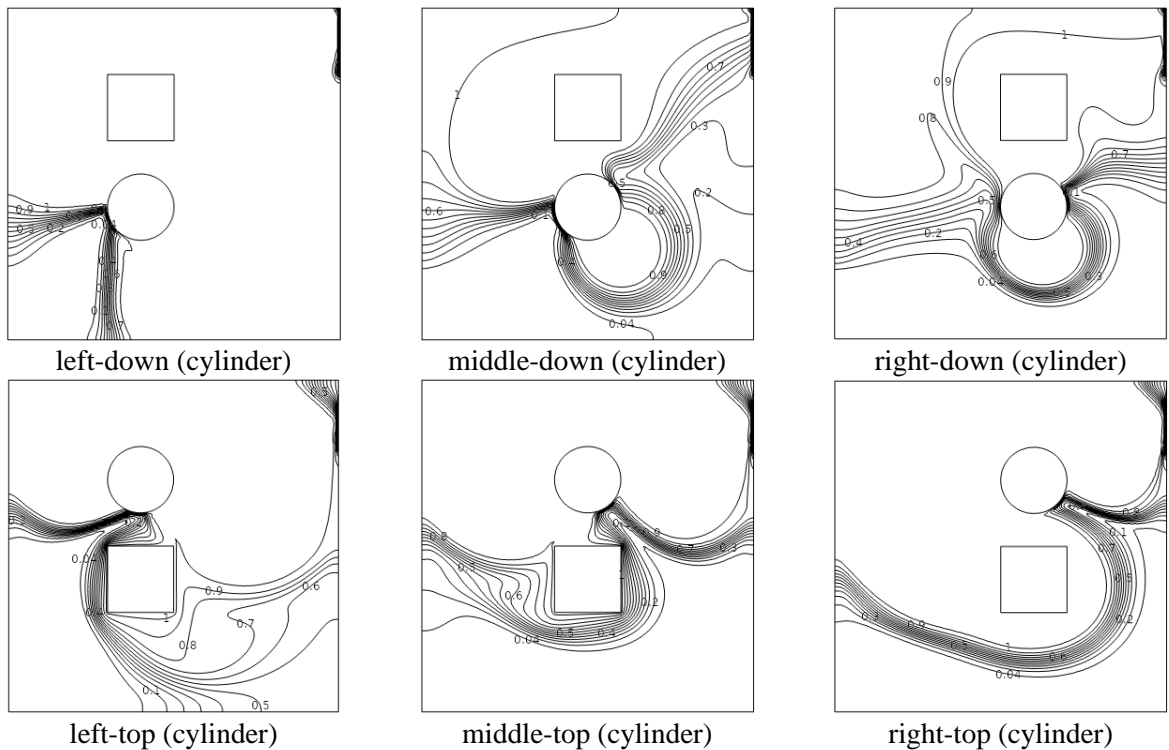
(**a**) Effect of horizontal position change on isothermal contours.

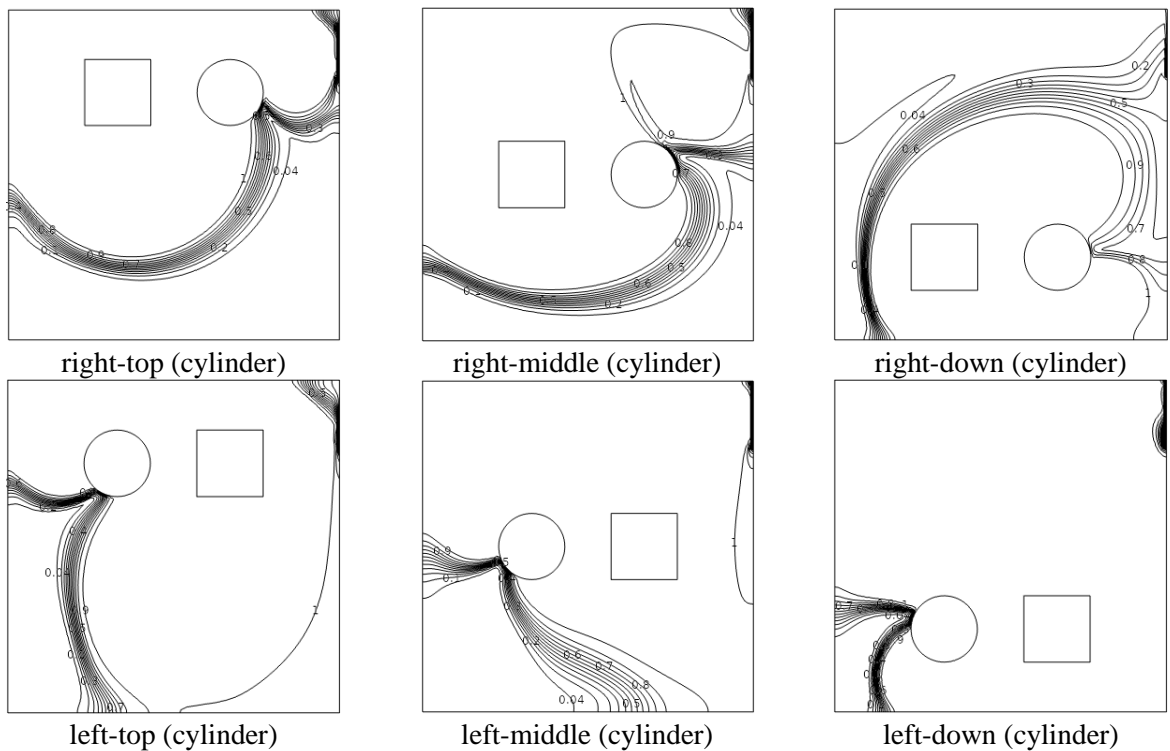
(**b**) Effect of vertical position change on isothermal contours.

**Fig. 21.** Variations of isothermal contours for the change of position of the obstacles at $\phi = 0.06, \omega = 20s^{-1}, R = 0.004, d_p = 5nm, Re = 100,$ and $Ri = 7$.

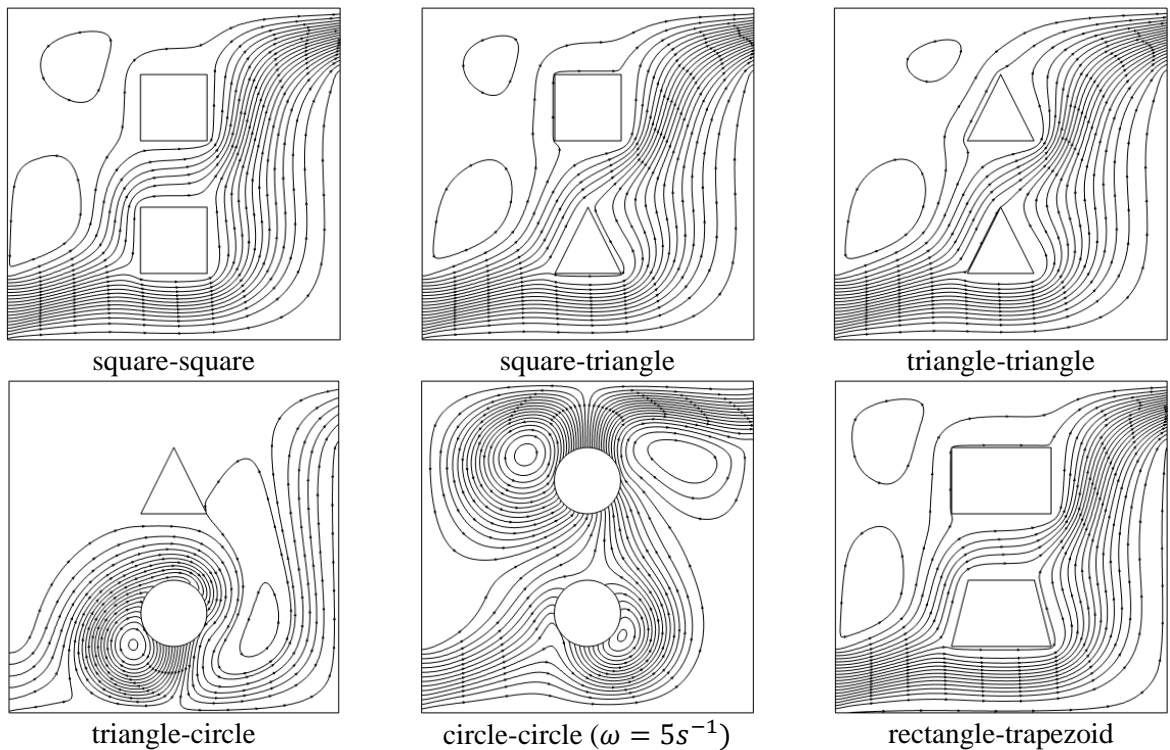
(**a**) Effect of change of shape of the obstacles on velocity streamlines.

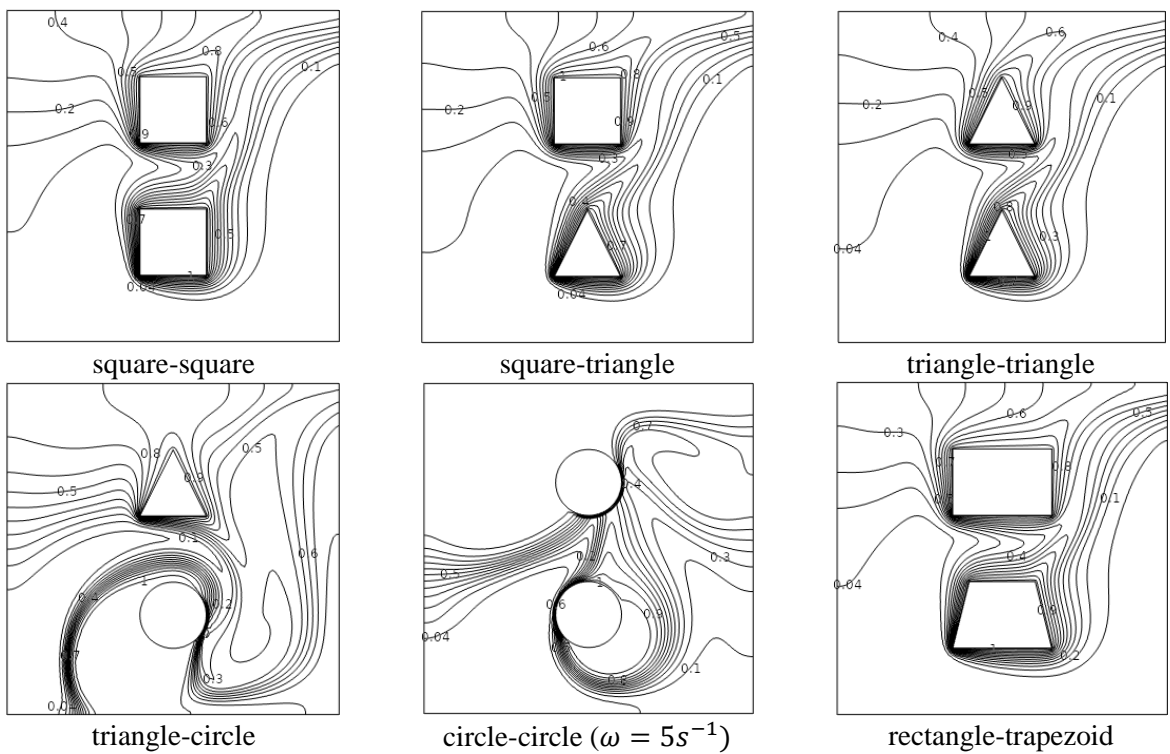
(**b**) Effect of change of shape of the obstacles on isothermal contours.

**Fig. 22.** Effects of change of shape of the obstacles on the velocity streamline and isothermal contours at $\phi = 0.06, R = 0.004, \omega = 20s^{-1}, d_p = 5nm, Re = 100,$ and $Ri = 7$.

## Competing interests
The authors declare no competing interests.

## Author Contributions
Hashnayne Ahmed: conceptualization, data curation, formal analysis, funding acquisition, investigation, methodology, resources, software, validation, visualization, writing – original draft, writing – review & editing. Chinmayee Podder: conceptualization, funding acquisition, methodology, project administration, supervision, validation, writing – review & editing.

## Data Availability
The data that support the findings of this study are available from the corresponding author upon reasonable request.

## Acknowledgments
This work is partially supported by the National Science & Technology (NST) fellowship program of the Ministry of Science and Technology, People's Republic of Bangladesh (Number – 2021/39.00.0000.012.05.20-05, Registration Number – 949).